\documentclass[10pt]{article}
\usepackage[T1]{fontenc}
\usepackage[utf8]{inputenc}
\usepackage{textcomp}
\usepackage{newtxtext,newtxmath}
\usepackage[letterpaper,margin=1in]{geometry}   
\usepackage{microtype}
\usepackage{setspace}
\usepackage[blocks]{authblk}

\usepackage{graphicx}
\usepackage{amsmath,mathtools}
\usepackage[version=4]{mhchem}
\usepackage{siunitx}
\usepackage{longtable,tabularx}
\usepackage{booktabs}
\usepackage[tableposition=top]{caption}
\usepackage{subcaption}
\usepackage{bm}
\usepackage{xcolor}
\usepackage[sort&compress,numbers]{natbib}

\usepackage[hyphens]{url}
\usepackage[colorlinks,allcolors=blue!50!black]{hyperref}
\usepackage[capitalise]{cleveref}
\crefname{appendix}{Appendix}{Appendices}
\newcommand{\refe}{\text{ref}}
\newcommand{\Pe}{\text{Pe}}
\newcommand{\Rey}{\text{Re}}
\newcommand{\Sc}{\text{Sc}}

\title{Nonpremixed Flames Near the Leading Edge of Parabolic Cylinders}
\author{Vinicius M. Sauer}
\author{Jose M. Martinez}
\affil{Department of Mechanical Engineering, California State University, Northridge, Northridge, CA 91330, USA}
\date{}
\begin{document}

\maketitle

\begin{abstract}
Nonpremixed flames in the forward stagnation region of porous cylinders are classically described by the first-order stagnation-point boundary layer, which neglects the effects of surface curvature and boundary-layer displacement. In the present work, the description is extended to second order for a porous parabolic cylinder using the method of matched asymptotic expansions. The flame is assumed to be a sheet detached from the porous wall with unity Lewis numbers, and the model accounts for thermal expansion and displacement of the entire reacting layer. The expansion is verified against laminar Navier-Stokes solutions of the same idealized problem, and in the resolved cases the two-term errors decrease approximately as the square of the expansion parameter. At Reynolds numbers typical of laboratory-scale burners, curvature and displacement move the flame away from the wall and reduce the scalar dissipation rate, wall shear, and wall heat flux, whereas the flame temperature changes little. These effects increase with injection, and displacement contributes more over a longer porous section. The second-order terms considerably reduce the errors of the first-order theory in the flame stand-off distance and the scalar dissipation rate.
\end{abstract}

\section*{Nomenclature}
{\singlespacing
\noindent\begin{longtable*}{@{}l @{\quad=\quad} l@{}}
    $A_q$, $\kappa_q$ & coefficient of the first-order correction of a flame quantity $q$ and its relative value, $A_q/q_1$ \\
    $\hat{a}$, $\hat{a}_\mathrm{eff}$ & nominal strain rate of the outer flow and effective strain rate at the stagnation line, \si{\per\second} \\
    $C$ & constant of the two-term error correlation \\
    $D_\infty$ & integral of $\theta - 1$ across the inner layer \\
    $E_{2,q}$, $e_{2,q}$ & correlated and computed relative two-term errors of a flame quantity $q$ \\
    $F$ & reduced stream function of the boundary-layer march \\
    $f$, $F_c$, $F_d$ & stream functions of the inner expansion \\
    $f_w$ & injection parameter, $-\rho_w v_w/(\varepsilon \sqrt{U_{1,1}})$ \\
    $g$, $G_c$, $G_d$ & mixture-fraction functions of the inner expansion \\
    $H$; $h$, $H_c$, $H_d$ & excess enthalpy; its functions of the inner expansion, normalized with $H_w$ \\
    $N$, $N_0$, $N_1$ & physical inner coordinate, $\sqrt{U_{1,1}}\, y/\varepsilon$, and its leading and first-order parts \\
    $N_u$ & numerical uncertainty \\
    $p$; $P_c$, $P_d$ & pressure, $(\hat{p} - \hat{p}_\infty)/(\hat{\rho}_\infty \hat{U}_\infty^2)$; its functions of the inner expansion \\
    $\Pe_R$, $\Pr$, $\Rey_R$ & P\'eclet, Prandtl and Reynolds numbers, with $\Pe_R = \Rey_R \Pr$ and $\Rey_R = \hat{a} \hat{R}^2/\hat{\nu}_\infty$ \\
    $q$ & flame quantity \\
    $\hat{R}$ & nose radius, \si{\milli\meter} \\
    $S$ & mass of oxidizer stream that reacts with a unit mass of fuel stream \\
    $T$, $\hat{T}$ & temperature rise, $(\hat{T} - \hat{T}_\infty)/\hat{T}_\infty$, and temperature, \si{\kelvin} \\
    $U_e$ & first-order slip velocity on the parabola, $\xi/(1 + \xi^2)^{1/2}$ \\
    $U_{i,j}$ & coefficient of $x^j$ in the slip velocity of order $i$ \\
    $\hat{U}_\infty$ & velocity scale, $\hat{a} \hat{R}$, \si{\centi\meter\per\second} \\
    $u$, $v$ & velocity components along and normal to the surface, scaled with $\hat{U}_\infty$ \\
    $x$, $y$ & coordinates along and normal to the surface, scaled with $\hat{R}$ \\
    $x_p$, $\xi_p$ & position of the edge of the porous section along the surface and in $\xi$ \\
    $\hat{x}_\mathrm{ax}$ & axial length of the porous section, \si{\milli\meter} \\
    $Y_F$, $Y_O$ & fuel and oxygen mass fractions, scaled with their injected and free-stream values \\
    $Z$ & mixture fraction \\
    $\beta_1$ & displacement constant of the first-order solution \\
    $\gamma$ & adiabatic temperature increment \\
    $\Delta$, $\tilde{\Delta}$ & displacement integral of the march and its remainder, $\Delta - \Delta_\infty$ \\
    $\delta$, $\delta^*$ & relative change of the strain rate due to displacement and displacement thickness \\
    $\varepsilon$ & expansion parameter, $\Rey_R^{-1/2}$ \\
    $\zeta$ & distance from the stagnation plane, scaled with $(\hat{\nu}_\infty/\hat{a})^{1/2}$ \\
    $\eta$, $\tilde{\eta}$ & Howarth-Dorodnitsyn inner variable and Görtler normal variable \\
    $\theta$, $\theta_1$ & temperature ratio, $\hat{T}/\hat{T}_\infty$, at leading order in the inner layer and its first-order correction \\
    $\mu$, $\rho$ & dynamic viscosity and density, scaled with $\hat{\mu}_\infty$ and $\hat{\rho}_\infty$ \\
    $\hat{\nu}$ & kinematic viscosity, \si{\meter\squared\per\second} \\
    $\xi$, $\eta_p$ & parabolic coordinates along and normal to the body \\
    $\hat{\tau}_w$ & wall shear stress, \si{\pascal} \\
    $\hat{\chi}$ & scalar dissipation rate, \si{\per\second} \\
    $\psi_w$ & stream function of the injected fluid \\
    \multicolumn{2}{@{}l}{\textbf{Subscripts}} \\
    $1$, $2$ & first- and second-order terms \\
    $c$, $d$ & curvature and displacement parts \\
    $f$, $s$ & flame and stoichiometric values \\
    $w$, $\infty$ & wall and free-stream values, or limit for $\eta$ or $\xi \rightarrow \infty$ \\
    \multicolumn{2}{@{}l}{\textbf{Superscripts}} \\
    $\hat{\phantom{a}}$ & dimensional quantity \\
    $'$ & derivative with respect to $\eta$, or to $\xi$ in \cref{app:U21} \\
\end{longtable*}}

\section{Introduction\label{sec:intro}}%
\begin{figure}[ht!]
    \centering
    \begin{subfigure}[b]{0.3\textwidth}
        \centering
        \includegraphics{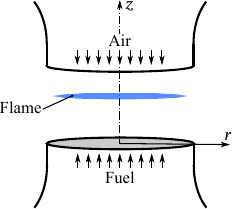}
        \caption{}
        \label{fig:counterflow_schem}
    \end{subfigure}
    \hspace{1ex}
    \begin{subfigure}[b]{0.3\textwidth}
        \centering
        \includegraphics{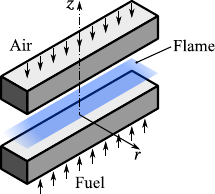}
        \caption{}
        \label{fig:slot_schem}
    \end{subfigure}
    \hspace{1ex}
    \begin{subfigure}[b]{0.3\textwidth}
        \centering
        \includegraphics{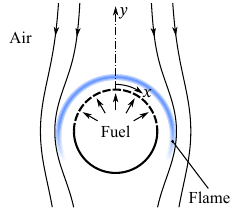}
        \caption{}
        \label{fig:tsuji_schem}
    \end{subfigure}
    \caption{Nonpremixed opposed-flow configurations. (a)~Axial axisymmetric flame; (b)~slot flame, with coordinates $z$ and $r$ normal and parallel to the stagnation plane; (c)~flame in the forward stagnation region of a porous cylinder, with $x$ and $y$ along and normal to the surface.}
    \label{fig:classical_schem}
\end{figure}

Counterflow and stagnation-point flames are among the most widely used configurations in combustion research. Their simple geometry reduces the flame near the stagnation line to a one-dimensional structure, allowing researchers to isolate and measure strain, extinction, flame structure, chemical kinetics, and transport. Since the analysis of the opposed-jet diffusion flame by Spalding \cite{spalding_theory_1961}, these flames (\cref{fig:counterflow_schem,fig:slot_schem}) have also served as the reference for the flamelet concept~\cite{peters_laminar_1984}. A classical variant places the flame in the forward stagnation region of a porous cylinder. In the burner of Tsuji and Yamaoka \cite{tsuji_counterflow_1967,tsuji_structure_1969,tsuji_structure_1971}, fuel is injected through the wall into an oxidizer stream, and a steady flame is embedded in the boundary layer near the stagnation line (\cref{fig:tsuji_schem}). Its structure follows from the boundary-layer equations, which Takeno \cite{takeno_gasdynamic_1972} extended to account for density variation across the layer and which Tsuji \cite{tsuji_counterflow_1982} reviewed in detail. The configuration was later used to compute the structure and extinction of methane-air flames with detailed chemistry \cite{dixon-lewis_calculation_1985} and to analyze extinction asymptotically \cite{krishnamurthy_asymptotic_1976}. Nonetheless, the configuration still raises new questions. Injection velocities comparable to the free-stream velocity blow the layer off the surface \cite{cole_blowhard_1968,li_aerodynamics_2022}. A porous cylinder in a planar counterflow forms the double Tsuji flame, a novel configuration introduced by Severino \cite{severino_diffusion_2020}, and further analyzed by Severino, Fachini, and coworkers \cite{severino_dynamics_2021,severino_mathematical_2022,fachini_double_2025,severino_double_2026,severino_counterflow_2026}, and by Li et al.~\cite{li_tsuji_2022}. At the other extreme, buoyant cylindrical and spherical porous burners were studied at low strain rates \cite{donini_buoyant_2020,tao_experimental_2022}, outside the forced, high-Reynolds-number regime of the present work.

The classical description of these flames is the first-order stagnation-point boundary layer with mass transfer, developed for high-speed flight \cite{fay_theory_1958} and solved for variable density by Libby \cite{libby_heat_1967} with the Howarth-Dorodnitsyn variable. Liu and Libby computed flames in this layer with finite-rate chemistry \cite{liu_boundary_1970} and in the flame-sheet limit \cite{liu_flame_1971}, with a porous-wall balance of the convective and diffusive fluxes of each species. The flame structure follows from ordinary differential equations, and the body enters only through the outer-flow strain rate. However, it is the leading term of an expansion for large Reynolds numbers, whereas laboratory burners typically operate at moderate ones. With the porous cylinder radius $\hat{R}$, the strain rate $\hat{a}$ and the free-stream kinematic viscosity $\hat{\nu}_\infty$, the Reynolds number $\Rey_R = \hat{a} \hat{R}^2/\hat{\nu}_\infty$ is about 4000 for $\hat{R} = \SI{25.4}{\milli\meter}$ at \SI{100}{\per\second}. The neglected terms are of relative order $\Rey_R^{-1/2}$, about \SI{1.6}{\percent}, multiplied by coefficients that are not known for a flame.

Second-order boundary-layer theory identifies these terms as the effects of surface curvature and displacement \cite{dyke_higher_1962a,dyke_higher_1962}. The displacement effect is global, since the slip velocity at the nose depends on the entire boundary-layer development, so it cannot be determined without the whole body \cite{dyke_higher_1964}. For a parabola in a uniform stream, both effects reduce the skin friction and heat transfer near the nose \cite{dyke_higher_1964}, and Navier-Stokes solutions of this body serve as benchmark results \cite{dennis_numerical_1971}. The theory was extended to compressible flow \cite{dyke_second_1962,davis_consistent_1970,papenfuss_secondorder_1976} and to suction and injection at constant density \cite{gersten_grenzschicht_1972,gersten_masstransfer_1973}, where injection increases all the second-order effects. In these analyses, the displacement coefficient was taken from the impermeable body or estimated from the local displacement thickness. In combustion, the two effects were studied separately. Kim et al.~\cite{kim_displacement_1993} showed that displacement from heat release modifies the strain imposed on counterflow flames, without curvature. Curvature effects on nonpremixed flames were studied mainly in tubular burners \cite{takeno_theoretical_1986,hu_experimental_2007,hu_structural_2009a}, and flamelet models with curvature contain an additional term of relative order $\Rey_R^{-1/2}$ for a layer on a body \cite{scholtissek_flamelet_2017}. In the Tsuji flame, heat release thickens the layer, the porous wall couples the mixture fraction to the flow, and the displacement depends on where the injection ends. However, to the authors' knowledge, the second-order inner problem with mass transfer and heat release has not been solved for a body with non-uniform curvature.

\begin{table}[t!]
    \caption{\label{tab:novelty} Formulation of Martinez and Sauer~\cite{martinez_nonpremixed_2025} and of the present work.}
    \centering
    \small
    \begin{tabular}{p{0.2\textwidth}p{0.3\textwidth}p{0.42\textwidth}}
        \toprule
        & Martinez and Sauer~\cite{martinez_nonpremixed_2025} & Present work \\
        \midrule
        Density & Constant & Variable, $\rho (1 + T) = 1$ and $\rho \mu = 1$ \\
        Wall condition for $Z$ & Dirichlet, $Z = 1$ & Flux balance at a porous wall (Robin) \\
        Temperature & Coupling function, Dirichlet wall & Excess enthalpy, prescribed wall temperature \\
        Displacement coefficient & $U_{2,1} = \num{-0.61}$ for every $f_w$ & Reacting march, porous section of finite extent \\
        Flame quantities & Temperature profile & Stand-off, temperature, and scalar dissipation corrections\\
        Comparison & Finite-rate computation in a finite gap & Navier-Stokes solutions of the same problem \\
        \bottomrule
    \end{tabular}
\end{table}

\begin{figure}[t!]
    \centering
    \includegraphics{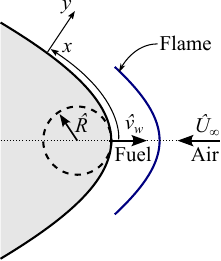}
    \caption{System representation, with the coordinates $x$ and $y$ along and normal to the surface, the nose radius $\bm{\hat{R}}$, the injection velocity $\bm{\hat{v}_w}$ and the free-stream velocity $\bm{\hat{U}_\infty}$.}
    \label{fig:system}
\end{figure}

This work addresses this gap for a porous parabolic cylinder (\cref{fig:system}), likely the simplest round-nosed body \cite{dyke_higher_1964}, whose outer potential flow is known in closed form and remains attached. It extends the constant-density formulation of Martinez and Sauer~\cite{martinez_nonpremixed_2025} -- \cref{tab:novelty} summarizes the differences. The curvature and displacement terms are classical \cite{dyke_higher_1962a,davis_consistent_1970,gersten_masstransfer_1973,papenfuss_secondorder_1976}, as is the displacement due to heat release \cite{kim_displacement_1993}. The new elements are the second-order inner problem with injection and heat release, the second-order mixture fraction with the porous-wall condition, and a displacement coefficient from the reacting layer of the whole body with a porous section of finite extent. The flame is a sheet detached from the wall, with unity Lewis numbers and thermal expansion. The analysis is restricted to the stagnation line, where the surface curvature variation enters at order $x^2$ and is undetermined, and the rest of the body enters only through the displacement coefficient. The expansion uses the nose radius, which differs from the radius of the stoichiometric surface by a relative amount of order $\Rey_R^{-1/2}$. The planar porous-wall flame at the same mass flux (i.e., the first-order solution) serves as the curvature reference, whereas a planar (slot) counterflow flame \cite{kee_computational_1989} provides context for the porous-wall effect. The expansion is verified against two-dimensional Navier-Stokes solutions of the same idealized problem to quantify the truncation error. The response of the flame to curvature, injection, and the extent of the porous section is then described at Reynolds numbers relevant to laboratory-scale burners. In the cases studied, thermal expansion increases the wall-shear curvature coefficient severalfold, and the displacement coefficient depends on injection along the entire porous section. \Cref{sec:model,sec:axi_models} describe the formulation and the finite-gap opposed flows. \Cref{sec:val,sec:results} present validation, numerical verification, flame response, and comparison with slot flames.

\section{Theoretical Model\label{sec:model}}%
In the coordinates of \cref{fig:system}, $x$ and $y$ are the distances along and normal to the surface, scaled with the nose radius $\hat{R}$, where the hat denotes dimensional quantities. The flame raises the temperature by about a factor of seven; hence, the variable-density formulation is the primary model. The Mach number is small, and the mean molecular weight is constant; hence $\rho (1 + T) = 1$, with density $\rho = \hat{\rho}/\hat{\rho}_\infty$ and temperature $T = (\hat{T} - \hat{T}_\infty)/\hat{T}_\infty$, where the subscript `$\infty$' denotes quantities in the air free stream. The viscosity follows $\rho \mu = 1$, with $\mu = \hat{\mu}/\hat{\mu}_\infty$. The specific heat and the Prandtl number are constant, and the Lewis numbers are unity \cite{libby_heat_1967}. The equations are written locally near the stagnation line using the vertex curvature, i.e., the metric $1 + y$, since the curvature variation enters only at relative order $x^2$. The conservation equations are given by \cite{goldstein_modern_1938,schlichting_boundarylayer_1979}
\begin{subequations}
    \begin{gather}
        \label{eq:mass}%
        \frac{\partial (\rho u)}{\partial x} +
        \frac{\partial}{\partial y}[(1 + y) \rho v] = 0
        \\
        \label{eq:long_mom}%
        \rho \left(
            \frac{u}{1 + y} \frac{\partial u}{\partial x} +
            v \frac{\partial u}{\partial y} +
            \frac{u v}{1 + y}
        \right) =
        - \frac{1}{1 + y} \frac{\partial p}{\partial x} +
        \frac{1}{\Rey_R} \left( \nabla \cdot \bm{\tau} \right)_x
        \\
        \label{eq:trans_mom}%
        \rho \left(
            \frac{u}{1 + y} \frac{\partial v}{\partial x} +
            v \frac{\partial v}{\partial y}
            - \frac{u^2}{1 + y}
        \right) =
        - \frac{\partial p}{\partial y} +
        \frac{1}{\Rey_R} \left( \nabla \cdot \bm{\tau} \right)_y
        \\
        \label{eq:energy}%
        \rho \left(
            \frac{u}{1 + y} \frac{\partial T}{\partial x} +
            v \frac{\partial T}{\partial y}
        \right) =
        \frac{1}{\Rey_R \Pr} \nabla \cdot \left( \mu \nabla T \right)
        - (1 + S) \gamma \dot{m}_F
        \\
        \label{eq:species}%
        \rho \left(
            \frac{u}{1 + y} \frac{\partial Y_i}{\partial x} +
            v \frac{\partial Y_i}{\partial y}
        \right) =
        \frac{1}{\Rey_R \Sc} \nabla \cdot \left( \mu \nabla Y_i \right)
        + S_i \dot{m}_F
    \end{gather}
\end{subequations}
where $\nabla \cdot (\mu \nabla \Phi) = (1 + y)^{-1} [\partial_x (\mu\, \partial_x \Phi/(1 + y)) + \partial_y ((1 + y) \mu\, \partial_y \Phi)]$ for a scalar $\Phi$ and $\bm{\tau}$ is the viscous stress tensor. The velocities $u$ and $v$ are scaled with $\hat{U}_\infty = \hat{a} \hat{R}$, where $\hat{a}$ is the nominal strain rate of the outer flow, and $p = (\hat{p} - \hat{p}_\infty)/\hat{\rho}_\infty \hat{U}_\infty^2$ is the scaled pressure. The mass fractions of fuel and oxygen are scaled with their values in the injected and free streams, $Y_F = \hat{Y}_F/\hat{Y}_{F,0}$ and $Y_O = \hat{Y}_O/\hat{Y}_{O,\infty}$, respectively. The parameter $S_i = s_i \hat{Y}_{F,0}/\hat{Y}_{i,\refe}$ is the mass of species $i$ that reacts with a unit mass of fuel stream \cite{linan_ignition_2015}, where $s_i$ is the mass of $i$ consumed per unit mass of fuel, which gives $S_F = 1$ and $S_O = S$. The fuel production rate is $\dot{m}_F = \hat{\dot{m}}_F \hat{R}/(\hat{\rho}_\infty \hat{U}_\infty \hat{Y}_{F,0})$, with $\hat{\dot{m}}_F$ in \si{\kilogram\per\cubic\meter\per\second}, and $\gamma = \hat{q} \hat{Y}_{F,0}/[\hat{c}_p \hat{T}_\infty (1 + S)]$ is the adiabatic temperature increment, where $\hat{q}$ is the heat of combustion and $\hat{c}_p$ the specific heat. The Reynolds, Prandtl, and Schmidt numbers are $\Rey_R = \hat{U}_\infty \hat{R}/\hat{\nu}_\infty$, $\Pr = \hat{\nu}/\hat{\alpha}$, and $\Sc = \hat{\nu}/\hat{\mathcal{D}} = \Pr$, where $\hat{\alpha}$ and $\hat{\mathcal{D}}$ are the thermal and mass diffusivities. Buoyancy, radiation, viscous dissipation, and Soret and Dufour effects are neglected.

An infinitely fast one-step reaction (Burke-Schumann limit) is assumed, which allows the energy and species equations to be replaced by the source-free equations for the mixture fraction $Z$ and the excess enthalpy $H$,
\begin{equation}\label{eq:coupling}%
    \rho \left(
        \frac{u}{1 + y} \frac{\partial \Phi}{\partial x} +
        v \frac{\partial \Phi}{\partial y}
    \right) =
    \frac{1}{\Pe_R} \nabla \cdot \left( \mu \nabla \Phi \right)
    \qquad\text{and}\qquad
    \Phi = Z, H
\end{equation}
where $\Pe_R = \Rey_R \Pr$ and
\begin{equation}\label{eq:ZH}%
    Z = \frac{S Y_{F} + 1 - Y_{O}}{S + 1}
    \qquad\text{and}\qquad
    H = T + \gamma \frac{S + 1}{S} \left( Y_O - 1 \right)
\end{equation}
The flame lies at the stoichiometric value $Z_s = 1/(S + 1)$, and the temperature and mass fractions follow from
\begin{equation}\label{eq:T_BS}%
    T = H + \frac{\gamma}{1 - Z_s} \min \left( 1, \frac{Z}{Z_s} \right)
    \qquad\text{and}\qquad
    Y_O = 1 - \frac{Z}{Z_s} \ (Z \leqslant Z_s)
    \qquad\text{and}\qquad
    Y_F = \frac{Z - Z_s}{1 - Z_s} \ (Z \geqslant Z_s)
\end{equation}
The porous wall -- indicated by the subscript `$w$' -- supplies a prescribed mass flux $\rho_w v_w$ of fuel stream, and its temperature is $T_w$. The species balances at a wall without surface reaction combine into a mixed (Robin) condition for $Z$ \cite{liu_boundary_1970,gersten_masstransfer_1973}. For a flame detached from the wall, $Y_{O,w} = 0$, and $H$ takes the value $H_w = T_w - \gamma/(1 - Z_s)$. Hence, $H = H_w h$, where $h$ is unity at the wall and vanishes in the free stream, and
\begin{subequations}\label{eq:bcs}%
    \begin{gather}
        u =
        (1 + y) \rho v - \rho_w v_w =
        \rho_w v_w \left( Z - 1 \right) - \frac{\mu}{\Pe_R} \frac{\partial Z}{\partial y} =
        h - 1 = 0, \ \text{at} \ y = 0
        \\
        u^2 + v^2 - 1 = p = Z = h = 0, \ \text{at } \ y \rightarrow \infty
    \end{gather}
\end{subequations}
A Dirichlet condition $Z = 1$ implies pure fuel at the surface, and it is recovered only under strong injection. The model is restricted to detached flames with the wall on the fuel side, $Z_w > Z_s$. For $Z_w < Z_s$, \cref{eq:T_BS} gives $Y_F = 0$ near the wall, and the fuel balance at the porous wall, $\rho_w v_w (Y_{F,w} - 1) = (\mu/\Pe_R)\, \partial Y_F/\partial y$, cannot be satisfied without a wall reaction.

\subsection{Outer Expansions\label{sec:outer}}
With the expansion parameter $\varepsilon = \Rey_R^{-1/2}$, the outer flow is expanded as $u = U_1 + \varepsilon U_2 + \mathcal{O}(\varepsilon^2)$, and similarly for $v$ and $p$. The flame lies inside the boundary layer, and the outer flows of both orders are irrotational and of uniform density. Near the stagnation point, the slip velocity is given by
\begin{equation}\label{eq:outer_series}%
    u =
    U_{1,1} x + U_{1,3} x^3 +
    \varepsilon \left[
        U_{2,1} x + U_{2,3} x^3
    \right] + \mathcal{O}(x^5) =
    U_{1,1} x \left[
        1 + \varepsilon \frac{U_{2,1}}{U_{1,1}} +
        \mathcal{O}(x^2)
    \right]
\end{equation}
where only odd powers of $x$ appear for symmetric flow \cite{schlichting_boundarylayer_1979}. The parabolic coordinates $(\xi, \eta_p)$ are defined by $\hat{x}_c + \mathrm{i}\hat{y}_c = \hat{R} [(\xi + \mathrm{i}\eta_p)^2 + 1]/2$, where $\hat{x}_c$ is the axial distance from the vertex and $\hat{y}_c$ the distance from the axis \cite{dyke_higher_1964}. The body lies at $\eta_p = 1$, and the first-order slip velocity is $U_e = \xi/(1 + \xi^2)^{1/2}$, which gives $U_{1,1} = 1$.

The second-order outer flow is driven by the displacement of the whole boundary layer. The injected mass flux is normal to the wall and uniform on $|x| < x_p$, i.e., $|\xi| < \xi_p$, where $x_p$ is the edge of the porous section, with $\rho_w v_w = -\varepsilon f_w$, where $f_w$ is the injection parameter. In the Görtler variables $\xi$ and $\tilde{\eta} = (1 + \xi^2)^{-1/2} \varepsilon^{-1} \int_0^y \rho \, \mathrm{d}y'$, with the stream function $\xi F(\xi, \tilde{\eta})$, the boundary-layer equations are
\begin{subequations}\label{eq:march}%
    \begin{gather}
        \partial_{\tilde{\eta}}^3 F + F\, \partial_{\tilde{\eta}}^2 F + \frac{\theta - (\partial_{\tilde{\eta}} F)^2}{1 + \xi^2} = \xi \left( \partial_{\tilde{\eta}} F\, \partial_\xi \partial_{\tilde{\eta}} F - \partial_{\tilde{\eta}}^2 F\, \partial_\xi F \right)
        \\
        \frac{1}{\Pr} \partial_{\tilde{\eta}}^2 \Phi + F\, \partial_{\tilde{\eta}} \Phi = \xi \left( \partial_{\tilde{\eta}} F\, \partial_\xi \Phi - \partial_{\tilde{\eta}} \Phi\, \partial_\xi F \right) \qquad\text{and}\qquad \Phi = Z, H
    \end{gather}
\end{subequations}
where $\theta = 1 + T$. At the wall, $F = \psi_w/\xi$, where $\psi_w = f_w \min(x, x_p)$ is the stream function of the injected fluid, and $\partial_{\tilde{\eta}} F = 0$. The Robin condition with the local mass flux is imposed for $Z$ on the porous section, and $\partial_{\tilde{\eta}} Z = 0$ is set beyond it. The wall value of $H$ is $T_w - \gamma \min(1, Z/Z_s)/(1 - Z_s)$, which allows oxygen to reach the impermeable wall downstream. The equations are marched in $\xi$ with a Keller box scheme, and the displacement thickness is $\delta^* = (1 + \xi^2)^{1/2} \Delta$, with
\begin{equation}\label{eq:Delta}%
    \Delta = \int_0^\infty \left( \theta - \partial_{\tilde{\eta}} F \right) \mathrm{d}\tilde{\eta}
\end{equation}
The $\xi \Delta_\infty$ part of the displacement, with Blasius value $\Delta_\infty = \num{1.2168}$, shifts the body to a confocal parabola without slip velocity. The remainder gives, through the Hilbert transform of a half-plane problem (\cref{app:U21}),
\begin{equation}\label{eq:U21}%
    U_{2,1} =
    - \frac{4}{\pi} \int_0^\infty \frac{1}{t} \frac{\mathrm{d} \tilde{\Delta}}{\mathrm{d} t} \mathrm{d}t
    + \frac{2}{\pi} f_w \left( \operatorname{arcsinh} \xi_p - \frac{\sqrt{1 + \xi_p^2}}{\xi_p} \right)
\end{equation}
where $\tilde{\Delta} = \Delta - \Delta_\infty$. For the impermeable parabola, $U_{2,1} = \num{-0.601}$, against $\num{-0.61}$ of Van Dyke \cite{dyke_higher_1964}. Earlier analyses with injection used Van Dyke's value \cite{gersten_masstransfer_1973} or $U_{2,1} \approx -\sqrt{U_{1,1}}\, \beta_1$, from the local displacement thickness near the nose \cite{gersten_grenzschicht_1972,papenfuss_grenzschichteffekte_1975}, where $\beta_1$ is defined later in \cref{eq:beta1}. We consider injection over an axial length $\hat{x}_\mathrm{ax} = \SI{25.4}{\milli\meter}$ from the vertex for $\hat{R} = \SI{12.7}{\milli\meter}$, \SI{25.4}{\milli\meter} and \SI{50.8}{\milli\meter}, i.e., a laboratory-scale geometry with $\hat{x}_\mathrm{ax}/\hat{R} = 2$, 1 and \num{0.5}. For these extents, $\xi_p = (2 \hat{x}_\mathrm{ax}/\hat{R})^{1/2}$, and $x_p = \num{2.958}$, \num{1.798}, and \num{1.148}, respectively. For injection limited to the porous section, no blow-off occurs within the expansion range.

\subsection{Inner Expansions}\label{sec:inner}
The inner region is described as~\cite{libby_heat_1967},
\begin{equation}\label{eq:eta}%
    \eta = \frac{\sqrt{U_{1,1}}}{\varepsilon} \int_0^y \rho \, \mathrm{d}t
    \qquad\text{and}\qquad
    y = \frac{\varepsilon}{\sqrt{U_{1,1}}} N(\eta)
    \qquad\text{and}\qquad
    N = \int_0^\eta (1 + T) \, \mathrm{d}t
\end{equation}
The expansion consists of the first-order function $f'$ (i.e., the Hiemenz solution with injection) and the curvature and displacement functions $F'_c$ and $F'_d$. Primes denote derivatives with respect to $\eta$, and the subscripts `$c$' and `$d$' identify the curvature and displacement parts of all variables,
\begin{subequations}
    \begin{gather}
        u = U_{1,1} x \left[
            f'(\eta) +
            \frac{\varepsilon}{\sqrt{U_{1,1}}} F'_c(\eta) +
            \varepsilon \frac{U_{2,1}}{U_{1,1}} F'_d(\eta)
        \right] +
        \mathcal{O}(\varepsilon^2 x, x^3)
        \\
        (1 + y) \rho v = - \varepsilon \sqrt{U_{1,1}} \left[
            f(\eta) +
            \frac{\varepsilon}{\sqrt{U_{1,1}}} F_c(\eta) +
            \varepsilon \frac{U_{2,1}}{U_{1,1}} F_d(\eta)
        \right] +
        \mathcal{O}(\varepsilon^3, \varepsilon x^2)
        \\
        p = \tfrac{1}{2}
        - \tfrac{1}{2} U_{1,1}^2 x^2 \left[
            1 +
            \frac{\varepsilon}{\sqrt{U_{1,1}}} P_c(\eta) +
            \varepsilon \frac{U_{2,1}}{U_{1,1}} P_d(\eta)
        \right] +
        \mathcal{O}(\varepsilon^2, x^4)
        \\
        Z = g(\eta) +
            \frac{\varepsilon}{\sqrt{U_{1,1}}} G_c(\eta) +
            \varepsilon \frac{U_{2,1}}{U_{1,1}} G_d(\eta) +
        \mathcal{O}(\varepsilon^2, x^2)
        \\
        H = H_w \left[
            h(\eta) +
            \frac{\varepsilon}{\sqrt{U_{1,1}}} H_c(\eta) +
            \varepsilon \frac{U_{2,1}}{U_{1,1}} H_d(\eta)
        \right] +
        \mathcal{O}(\varepsilon^2, x^2)
    \end{gather}
\end{subequations}
where the remainders include the longitudinal orders. The terms of relative order $x^2$, which contain the variation of the surface curvature and $U_{1,3}$, are not determined. The temperature is $1 + T = \theta + \varepsilon \theta_1$, with $\theta_1 = \theta_c/\sqrt{U_{1,1}} + (U_{2,1}/U_{1,1})\, \theta_d$, and hence $N = N_0 + \varepsilon N_1$, with $N_0 = \int_0^\eta \theta\, \mathrm{d}\eta'$ and $N_1 = \int_0^\eta \theta_1\, \mathrm{d}\eta'$. Substituting into \cref{eq:mass,eq:long_mom,eq:trans_mom} gives
\begin{subequations}\label{eq:inner_flow}%
    \begin{gather}
        \label{eq:f}%
        f''' + ff'' + \theta - f'^2 = 0
        \\
        \label{eq:Fd}%
        F'''_d + f F''_d - 2 f' F'_d + f'' F_d + \theta_d
        = - \theta P_d
        \\
        \label{eq:Fc}%
        F'''_c + f F''_c - 2 f' F'_c + f'' F_c + \theta_c
        = - \theta P_c - \theta f f' - N_0 f''' - \theta f'' + \theta' f'
    \end{gather}
\end{subequations}
where $P_d = 2$ and
\begin{equation}\label{eq:Pc}%
    P_c = -2 \left( \eta + D_\infty + J \right)
    \qquad\text{and}\qquad
    D_\infty = \int_0^\infty (\theta - 1) \, \mathrm{d}\eta
    \qquad\text{and}\qquad
    J = \int_\eta^\infty \left( 1 - f'^2 \right) \mathrm{d}\eta'
\end{equation}
from the normal momentum equation, $P'_c = -2 f'^2$, and matching \cite{davis_consistent_1970}. For $\theta = 1$, \cref{eq:inner_flow} reduces to the equations of Gersten and Gross \cite{gersten_masstransfer_1973}, and \cref{eq:Fc} agrees with Papenfuss \cite{papenfuss_secondorder_1976} for a viscosity proportional to temperature. The temperature couples the flow to the scalars through $\theta = 1 + H_w h + \gamma \min(1, g/Z_s)/(1 - Z_s)$ and $\theta_{c,d} = (\partial T/\partial Z) G_{c,d} + H_w H_{c,d}$, where $\partial T/\partial Z = \gamma/[Z_s (1 - Z_s)]$ on the oxidizer side and zero on the fuel side. The scalar equations are
\begin{subequations}\label{eq:inner_z}%
    \begin{gather}
        \frac{1}{\Pr} g'' + f g' = 0
        \qquad\text{and}\qquad
        \frac{1}{\Pr} G_d'' + f G_d' + F_d g' = 0
        \\
        \frac{1}{\Pr} G_c'' + f G_c' + F_c g' + \frac{1}{\Pr} \left( \theta g' + N_0 g'' \right) = 0
    \end{gather}
\end{subequations}
and the same equations hold for $h$, $H_c$ and $H_d$. The boundary conditions are
\begin{subequations}\label{eq:inner_bcs}%
    \begin{gather}
        f - f_w = f' = F_c = F'_c = F_d = F'_d = 0, \ \text{at } \ \eta = 0
        \\
        g' + \Pr f_w (g - 1) = G'_c + \Pr f_w G_c = G'_d + \Pr f_w G_d = h - 1 = H_c = H_d = 0, \ \text{at } \ \eta = 0
        \\
        f' - 1 = F''_c + 1 = F'_d - 1 = g = G_c = G_d = h = H_c = H_d = 0, \ \text{at } \ \eta \rightarrow \infty
    \end{gather}
\end{subequations}
where $f_w = -\rho_w v_w/(\varepsilon \sqrt{U_{1,1}})$ is based on the injected mass flux \cite{libby_heat_1967,gersten_masstransfer_1973}. Matching gives $F'_c \sim -(\eta + D_\infty)$ and the outer transpiration $V_2(x,0) = \sqrt{U_{1,1}} (\beta_1 + D_\infty)$, with
\begin{equation}\label{eq:beta1}%
    \beta_1 = \lim_{\eta \rightarrow \infty} (\eta - f)
\end{equation}
The constant-density model is recovered for $\theta = 1$, $\theta_{c,d} = 0$, $N = \eta$ and $D_\infty = 0$. In this limit, $g = g_w \int_\eta^\infty E\, \mathrm{d}\eta' / I_E$, with $E = \exp(-\Pr \int_0^\eta f\, \mathrm{d}\eta')$, $I_E = \int_0^\infty E\, \mathrm{d}\eta$ and the wall value $g_w = -\Pr f_w I_E/(1 - \Pr f_w I_E)$ \cite{gersten_masstransfer_1973}.

The slip velocity $\varepsilon U_{2,1} x$ changes the strain to $U_{1,1}(1 + \delta)$, with $\delta = \varepsilon U_{2,1}/U_{1,1}$. Rescaling the first-order solution at a fixed injected mass flux, $\eta \rightarrow \eta \sqrt{1 + \delta}$ and $f_w \rightarrow f_w/\sqrt{1 + \delta}$, gives
\begin{equation}\label{eq:Fd_strain}%
    F_d = \frac{1}{2} \left( f + \eta f' - f_w \frac{\partial f}{\partial f_w} \right)
    \qquad\text{and}\qquad
    G_d = \frac{1}{2} \left( \eta g' - f_w \frac{\partial g}{\partial f_w} \right)
\end{equation}
Therefore, at the stagnation line, displacement changes the effective strain rate to $\hat{a}_\mathrm{eff} = \hat{a} (1 + \varepsilon U_{2,1}/U_{1,1})$. All the Reynolds numbers below are based on the nominal strain rate. At fixed $f_w$, the mass flux $\hat{\rho}_w \hat{v}_w = -f_w (\hat{\rho}_\infty \hat{\mu}_\infty \hat{a})^{1/2}$ grows as $\hat{a}^{1/2}$, and hence a sweep of $\hat{a}$ at constant $f_w$ does not keep the injection fixed.

\subsection{Flame Quantities and Accuracy}\label{sec:flame2}
The flame sheet lies at $\eta_f = \eta_{f1} + \varepsilon \eta_{f2}$, where $g(\eta_{f1}) = Z_s$ and
\begin{equation}\label{eq:etaf2}%
    \eta_{f2} = - \frac{1}{g'(\eta_{f1})} \left[ \frac{G_c(\eta_{f1})}{\sqrt{U_{1,1}}} + \frac{U_{2,1}}{U_{1,1}} G_d(\eta_{f1}) \right]
\end{equation}
The physical stand-off distance is $y_f = \varepsilon N(\eta_f)/\sqrt{U_{1,1}}$, and expanding about $\eta_{f1}$ with $\partial_\eta N_0 = \theta$ gives
\begin{equation}\label{eq:yf2}%
    \frac{\sqrt{U_{1,1}}}{\varepsilon}\, y_f =
    N_0(\eta_{f1}) + \varepsilon \left[ \theta(\eta_{f1})\, \eta_{f2} + N_1(\eta_{f1}) \right]
\end{equation}
The flame temperature $T_f$ and the stoichiometric scalar dissipation rate $\hat{\chi}_s$, i.e., the value of $\hat{\chi} = 2 \hat{\mathcal{D}} |\nabla Z|^2$ at $Z = Z_s$, are
\begin{gather}
    \label{eq:Tf2}%
    T_f = \frac{\gamma}{1 - Z_s} + H_w \left\{ h(\eta_{f1}) + \varepsilon \left[ h'(\eta_{f1}) \eta_{f2} + \frac{H_c(\eta_{f1})}{\sqrt{U_{1,1}}} + \frac{U_{2,1}}{U_{1,1}} H_d(\eta_{f1}) \right] \right\}
    \\
    \label{eq:chi2}%
    \frac{\hat{\chi}_s}{\hat{a}} = \frac{2}{\Pr} g'^2(\eta_{f1}) + \varepsilon \frac{4}{\Pr} g'(\eta_{f1}) \left[ g''(\eta_{f1}) \eta_{f2} + \frac{G'_c(\eta_{f1})}{\sqrt{U_{1,1}}} + \frac{U_{2,1}}{U_{1,1}} G'_d(\eta_{f1}) \right]
\end{gather}
where the density factors cancel due to $\rho \mu = 1$ (\cref{app:inner}). The leading coefficient in $x$ of the wall shear stress is proportional to $f''(0) + \varepsilon [F''_c(0)/\sqrt{U_{1,1}} + (U_{2,1}/U_{1,1}) F''_d(0)]$.

Each quantity is written as $q = q_1 + \varepsilon A_q + \mathcal{O}(\varepsilon^2)$, with $A_q = q_{2,c}/\sqrt{U_{1,1}} + (U_{2,1}/U_{1,1})\, q_{2,d}$, and its relative first-order correction is $\varepsilon \kappa_q$, with $\kappa_q = A_q/q_1$. The first-order model is considered adequate when the total corrections are below an engineering tolerance of \SI{10}{\percent},
\begin{equation}\label{eq:first_order}%
    \varepsilon |\kappa_{y_f}| \leqslant 0.1
    \qquad \text{and} \qquad
    \varepsilon |\kappa_\tau| \leqslant 0.1
    \qquad \text{and} \qquad
    \varepsilon |\kappa_\chi| \leqslant 0.1
\end{equation}
for the stand-off distance, the wall shear, and the scalar dissipation rate, respectively. However, this tolerance does not define a range of validity of the expansion. The two-term error is of order $\varepsilon^2$, and its coefficient requires solving the third-order problem. Therefore, it is described by the empirical correlation
\begin{equation}\label{eq:validity}%
    E_{2,q} = C \left( \varepsilon \kappa_q \right)^2
    \qquad \text{and} \qquad
    \kappa_q = \frac{1}{q_1} \left( \frac{q_{2,c}}{\sqrt{U_{1,1}}} + \frac{U_{2,1}}{U_{1,1}} q_{2,d} \right)
\end{equation}
where $C$ is a calibration constant. The relation between the third-order term and $A_q^2$ does not follow from the asymptotic ordering. Therefore, \cref{eq:validity} is a correlation for the cases evaluated. The two-term expansion is considered accurate when $E_{2,q} \leqslant \SI{2}{\percent}$ for the stand-off distance and $\SI{5}{\percent}$ for the scalar dissipation rate. The tolerance is larger for the scalar dissipation rate, since it depends on the square of the mixture-fraction gradient, which approximately doubles its relative error. Both tolerances exceed the numerical uncertainty of the Navier-Stokes solutions (\cref{sec:cfd}). They are also small compared with the differences between the flame sheet and the detailed finite-rate formulation (\cref{sec:slot}). The correlation is used only where the first-order flame lies more than one viscous length, $(\hat{\nu}_\infty/\hat{a})^{1/2}$, from the wall, i.e., for $f_w \lesssim -0.3$. The boundary-layer structure also requires $-f_w \varepsilon \lesssim 0.3$.

The numerical solutions in \cref{sec:cfd} solve the same idealized equations and therefore measure the truncation error. The combined modeling differences against a detailed finite-rate and transport formulation are evaluated for selected planar cases in \cref{sec:slot}. The total error relative to experimental methane flames, which also includes radiation and buoyancy, is not established. The buoyancy parameter $\hat{g}/(\hat{a}^2 \hat{R})$, where $\hat{g}$ is the gravitational acceleration, is about 0.17 at $\Rey_R = 2000$ for $\hat{R} = \SI{25.4}{\milli\meter}$, and its effect on the nose region is assumed to be small. At a fixed $\Rey_R$, it grows as $\hat{R}^3$, and hence it is close to one for $\hat{R} = \SI{50.8}{\milli\meter}$.

\section{Finite-Gap Opposed Flows}\label{sec:axi_models}%
The axial and slot opposed flows of \cref{fig:counterflow_schem,fig:slot_schem} are described in a finite gap by one-dimensional equations \cite{kee_chemically_2003}. With $\hat{z}$ the coordinate normal to the stagnation surface, $\hat{r}$ the coordinate parallel to it, and $\hat{u}_z$ and $\hat{u}_r$ the corresponding velocities,
\begin{subequations}
    \begin{gather}
        \label{eq:gf_mass}%
        \frac{\mathrm{d}}{\mathrm{d}\hat{z}} \left(
            \hat{\rho} \hat{u}_z
        \right) +
        2^{b} \hat{\rho} \hat{W} = 0
        \\
        \label{eq:gf_mom}%
        \hat{\rho} \hat{u}_z \frac{\mathrm{d}\hat{W}}{\mathrm{d}\hat{z}} +
        \hat{\rho} \hat{W}^{2} -
        \frac{\mathrm{d}}{\mathrm{d}\hat{z}} \left(
            \hat{\mu} \frac{\mathrm{d}\hat{W}}{\mathrm{d}\hat{z}}
        \right) =
        - \frac{1}{\hat{r}} \frac{\partial \hat{p}}{\partial \hat{r}} =
        - \hat{\Lambda}
        \\
        \hat{\rho} \hat{u}_z \frac{\mathrm{d}\hat{Y}_{i}}{\mathrm{d}\hat{z}} =
        - \frac{\mathrm{d}}{\mathrm{d}\hat{z}} \left( \hat{\rho} \hat{\mathcal{V}}_{i} \hat{Y}_{i} \right) +
        \hat{\dot{m}}_{i}
        \label{eq:gf_sc}%
        \\
        \label{eq:gf_ec}%
        \hat{\rho} \hat{u}_z \hat{c}_{p} \frac{\mathrm{d}\hat{T}}{\mathrm{d}\hat{z}} =
        \frac{\mathrm{d}}{\mathrm{d}\hat{z}} \left(
        \hat{\lambda} \frac{\mathrm{d}\hat{T}}{\mathrm{d}\hat{z}}
        \right) -
        \sum_{i = 1}^{N_\mathrm{sp}}{
            \hat{\rho} \hat{c}_{p,i} \hat{\mathcal{V}}_{i} \hat{Y}_{i} \frac{\mathrm{d}\hat{T}}{\mathrm{d}\hat{z}}
        } -
        \sum_{i = 1}^{N_\mathrm{sp}}{
            \hat{h}_{i} \hat{\dot{m}}_{i}
        }
    \end{gather}
\end{subequations}
where $\hat{W} = \hat{u}_r/\hat{r}$, $\hat{Y}_i$ is the physical mass fraction, $\hat{\mathcal{V}}_i$ the diffusion velocity, $\hat{\dot{m}}_i$ the volumetric production rate, $\hat{c}_{p,i}$ and $\hat{h}_i$ the specific heat and enthalpy of species $i$, $\hat{\lambda}$ the thermal conductivity, and $\hat{\Lambda}$ the pressure curvature eigenvalue. The parameter $b$ is one and zero for axisymmetric and slot flows, respectively. Hence, an axial solver can be extended to slot flow by changing the factor in the continuity equation. The slot solutions use a modified Cantera code based on v3.1.0 \cite{goodwin_cantera_2024}, whose repository is given in the Supplemental Material. This implementation is verified against the Case B measurements, and computations of Sung et al.~\cite{sung_structural_1995} and against a separately implemented planar solution, as shown in the Supplemental Material. The finite-rate porous-wall flame is computed with the same code as a planar stagnation flow at first order. The porous wall is a no-slip fuel inlet at $\hat{T}_w = \SI{300}{\kelvin}$, with the species flux conditions of the porous wall and the mass flux $\hat{\rho}_w \hat{v}_w = -f_w (\hat{\rho}_\infty \hat{\mu}_\infty \hat{a})^{1/2}$. The oxidizer enters as the Hiemenz flow of strain rate $\hat{a}$. The extinction strain is a refined numerical estimate of the upper burning-branch extinction strain. The prescribed control strain is raised until the burning solution is lost, and each cold state is rematched to the prescribed strain. The bracket between the last burning and the first cold control strain (i.e., the search resolution) is below \SI{0.4}{\percent}. In addition, a finer grid changes the estimate by less than \SI{0.7}{\percent}, and doubling the domain width changes the results by less than \SI{0.05}{\percent}. The detailed flame is located where the Bilger mixture fraction, normalized with the supplied fuel and oxidizer streams, equals $Z_s$. At $f_w = -0.8$ and $\hat{a} = \SI{100}{\per\second}$, this surface lies about \SI{3}{\percent} closer to the wall than the peak temperature and the peak heat release. The extinction values and selected comparisons with the flame sheet are listed in the Supplemental Material. All finite-rate solutions use the San Diego mechanism \cite{ucsd_san_2016} and mixture-averaged transport. In the slot, the strain rate is defined from the eigenvalue as $\hat{a} = (-\hat{\Lambda}/\hat{\rho}_\infty)^{1/2}$, where $\hat{\rho}_\infty$ is the oxidizer density, and it equals the tangential velocity gradient of an inviscid stagnation flow of the oxidizer with the same pressure field.

\section{Validation}\label{sec:val}%
\begin{table}[t!]
    \caption{\label{tab:validation} Wall values of the inner functions for $\bm{\Pr = 0.7}$, $\bm{Z_s = 0.2278}$, $\bm{\gamma = 5.760}$ and $\bm{T_w = 0}$. (a)~Flow; (b)~mixture fraction and normalized excess enthalpy. The values do not depend on $\bm{\Rey_R}$, and $\bm{f_w < 0}$ denotes injection.}
    \centering
    \small
    \setlength{\tabcolsep}{3pt}
    \begin{tabular}{lrrrrrrrrrr}
        \toprule
        \multicolumn{11}{l}{(a)~Flow} \\
        \midrule
         & \multicolumn{5}{c}{Constant density} & \multicolumn{5}{c}{Variable density} \\
        \cmidrule(lr){2-6} \cmidrule(lr){7-11}
        $f_w$ & $f''(0)$ & $\beta_1$ & $D_\infty$ & $F''_c(0)$ & $F''_d(0)$ & $f''(0)$ & $\beta_1$ & $D_\infty$ & $F''_c(0)$ & $F''_d(0)$ \\
        \midrule
        \num{0.0} & \num{1.2326} & \num{0.6479} & \num{0.0000} & \num{-1.9133} & \num{1.8489} & \num{1.2326} & \num{0.6479} & \num{0.0000} & \num{-1.9133} & \num{1.8489} \\
        \num{-0.5} & \num{0.9692} & \num{1.2810} & \num{0.0000} & \num{-1.8883} & \num{1.5731} & \num{1.9523} & \num{0.0949} & \num{4.5003} & \num{-11.1828} & \num{3.0175} \\
        \num{-1.0} & \num{0.7566} & \num{1.9450} & \num{0.0000} & \num{-1.8545} & \num{1.3217} & \num{1.5156} & \num{0.1305} & \num{6.9566} & \num{-13.5522} & \num{2.8023} \\
        \num{-2.0} & \num{0.4758} & \num{3.3616} & \num{0.0000} & \num{-1.7672} & \num{0.9135} & \num{0.6298} & \num{0.9956} & \num{8.9707} & \num{-8.2967} & \num{1.5301} \\
        \bottomrule
        \end{tabular}
        \par\medskip
        \begin{tabular}{lrrrrrrrrrrrr}
        \toprule
        \multicolumn{13}{l}{(b)~Scalars} \\
        \midrule
         & \multicolumn{6}{c}{Constant density} & \multicolumn{6}{c}{Variable density} \\
        \cmidrule(lr){2-7} \cmidrule(lr){8-13}
        $f_w$ & $g(0)$ & $G_c(0)$ & $G_d(0)$ & $h'(0)$ & $H'_c(0)$ & $H'_d(0)$ & $g(0)$ & $G_c(0)$ & $G_d(0)$ & $h'(0)$ & $H'_c(0)$ & $H'_d(0)$ \\
        \midrule
        \num{0.0} & \num{0.0000} & \num{0.0000} & \num{0.0000} & \num{-0.4959} & \num{0.1281} & \num{-0.2479} & \num{0.0000} & \num{0.0000} & \num{0.0000} & \num{-0.4959} & \num{0.1281} & \num{-0.2479} \\
        \num{-0.5} & \num{0.5441} & \num{0.1227} & \num{-0.1988} & \num{-0.2933} & \num{0.1451} & \num{-0.2351} & \num{0.4499} & \num{0.3319} & \num{-0.1599} & \num{-0.4280} & \num{0.5739} & \num{-0.2765} \\
        \num{-1.0} & \num{0.8278} & \num{0.1380} & \num{-0.1865} & \num{-0.1456} & \num{0.1410} & \num{-0.1905} & \num{0.7146} & \num{0.7363} & \num{-0.2114} & \num{-0.2795} & \num{1.0092} & \num{-0.2898} \\
        \num{-2.0} & \num{0.9882} & \num{0.0401} & \num{-0.0382} & \num{-0.0167} & \num{0.0575} & \num{-0.0547} & \num{0.9633} & \num{0.4848} & \num{-0.1043} & \num{-0.0533} & \num{0.7314} & \num{-0.1574} \\
        \bottomrule
    \end{tabular}
\end{table}

\subsection{Parabolic Cylinders}\label{sec:parabola_val}%
The inner problem is solved with the collocation method \texttt{solve\_bvp} of SciPy (Python 3.13) \cite{kierzenka_bvp_2001,virtanen_scipy_2020}. For variable density, the domain is split at the flame sheet, with $\eta_{f1}$ as an unknown parameter (\cref{app:inner}). With $\eta \leqslant 16$ and a tolerance of \num{e-8}, the results change by less than \num{e-8} with $\eta \leqslant 24$. The reference mixture is that of Hu et al.~\cite{hu_structural_2009a}, \SI{30}{\percent} \ce{CH4} in \ce{N2} against air with \SI{1}{\percent} \ce{Ar} (mole fractions), both at \SI{300}{\kelvin}. Its kinematic viscosity, $\hat{\nu}_\infty = \SI{1.588e-5}{\meter\squared\per\second}$ from Cantera with mixture-averaged transport, is used for every conversion between $\Rey_R$ and $\hat{a}$. The San Diego mechanism \cite{ucsd_san_2016} gives $S = \num{3.389}$ and $Z_s = \num{0.2278}$. With complete combustion to \ce{CO2} and \ce{H2O}, $\gamma = \num{5.760}$, and we set $T_w = 0$ and $\Pr = \num{0.7}$. Since $g$ and $h$ obey the same equation and vanish outside the layer, $g = g_w h$, and the first-order flame temperature is $T_{f1} = \gamma/(1 - Z_s) + H_w Z_s/g_w$. It falls as $g_w$ approaches $Z_s$, i.e., as the flame approaches the cold wall.

\Cref{tab:validation} lists the wall values of the inner functions. For $f_w = 0$, there is no flame in the layer, $f''(0)$ is the Hiemenz value, and $F''_d(0) = \tfrac{3}{2} f''(0)$, as \cref{eq:Fd_strain} requires. The constant-density rows agree with Gersten and Gross \cite{gersten_masstransfer_1973}, Libby \cite{libby_heat_1967}, and Gersten et al.~\cite{gersten_grenzschicht_1972}. Without the flame, the variable-density equations reproduce the cold-wall results of Libby, $f''(0) = \num{0.6071}$ and $\beta_1 = \num{1.0560}$ for $T_w = -1$, which is a mathematical limit used only as a benchmark. They also reproduce the compressible coefficients of Papenfuss \cite{papenfuss_grenzschichteffekte_1975} for a wall temperature twice the free-stream value, $f''(0) = \num{1.767603}$, $F''_c(0) = \num{-4.129191}$ and $F''_d(0) = \num{2.651405}$, which verifies the second-order density in \cref{eq:Fc}. For the impermeable parabola, the wall shear and heat transfer are proportional to $\num{1.2326} - \num{3.025} \varepsilon$ and $\num{0.4959} - \num{0.277} \varepsilon$, against $\num{1.2326} - \num{3.03} \varepsilon$ and $\num{0.4959} - \num{0.279} \varepsilon$ of Van Dyke \cite{dyke_higher_1964}. Thermal expansion doubles the wall shear for $f_w = -1$ and increases $F''_c(0)$ by a factor of about seven.

\begin{figure}[ht!]
    \centering
    \includegraphics{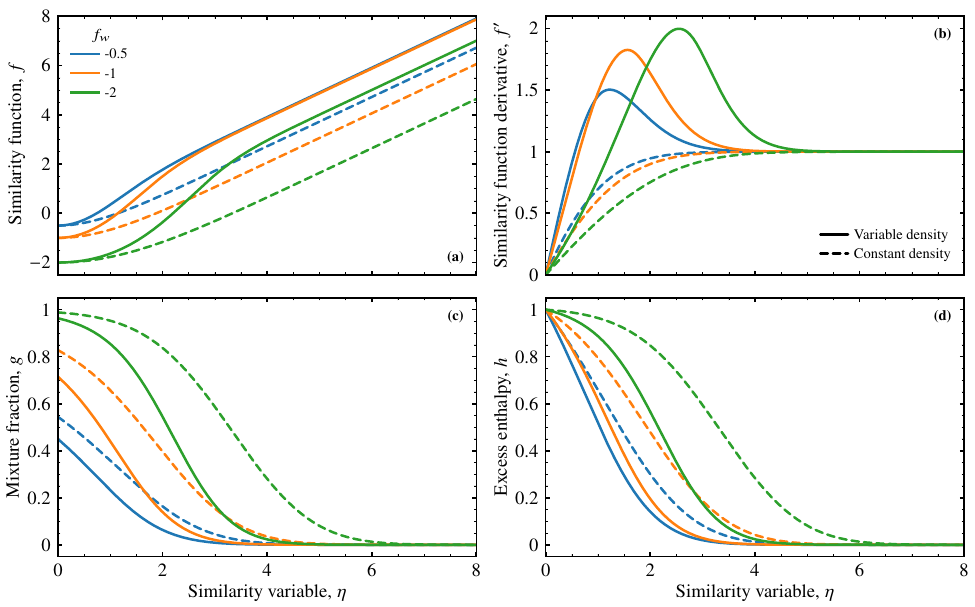}
    \caption{First-order inner solutions for $\bm{f_w = -0.5}$, $\bm{-1}$ and $\bm{-2}$, $\bm{Z_s = 0.228}$ and $\bm{\Pr = 0.7}$. (a)~Stream function $\bm{f}$; (b)~longitudinal velocity $\bm{f'}$; (c)~mixture fraction $\bm{g}$; (d)~normalized excess enthalpy $\bm{h}$. Solid lines: variable density; dashed lines: constant density.}
    \label{fig:inner_first}
\end{figure}

\begin{figure}[ht!]
    \centering
    \includegraphics{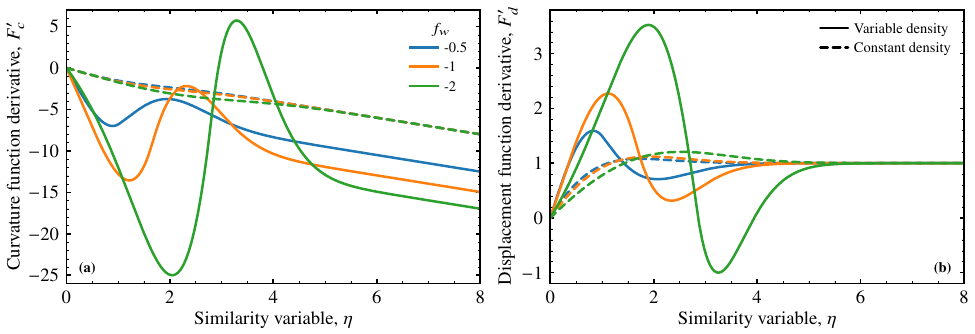}
    \caption{Second-order longitudinal velocity functions for $\bm{f_w = -0.5}$, $\bm{-1}$ and $\bm{-2}$ and $\bm{Z_s = 0.228}$. (a)~Curvature, $\bm{F'_c}$; (b)~displacement, $\bm{F'_d}$. Solid lines: variable density; dashed lines: constant density.}
    \label{fig:inner_second}
\end{figure}

\Cref{fig:inner_first} shows the first-order functions. With variable density, $f'$ exceeds the outer value by a factor of up to two near the flame, due to the acceleration of the hot gas by the same pressure gradient. The physical stand-off distance is larger than for constant density, e.g., $N_0(\eta_{f1}) = \num{5.56}$ against $\eta_{f1} = \num{2.65}$ at $f_w = -1$. Similarly, the second-order functions in \cref{fig:inner_second} change sharply near the flame, where $\theta_{c,d}$ concentrates.

\subsection{Constant-Density Limit}\label{sec:cd_limit}
The constant-density limit is verified against incompressible Navier-Stokes solutions with passive scalars, using the grids in \cref{sec:cfd}. For $f_w = -1$ and $-2$ and $\Rey_R = 250$ to \num{1.5e4}, the two-term errors decrease approximately as $\varepsilon^{1.75}$ to $\varepsilon^{2}$. The slip-velocity gradient extrapolated above the layer is $\num{-0.600}$ for the impermeable wall, compared with $\num{-0.601}$ from the march.

\section{Results}\label{sec:results}
\subsection{Displacement Coefficient}
\begin{figure}[ht!]
    \centering
    \includegraphics{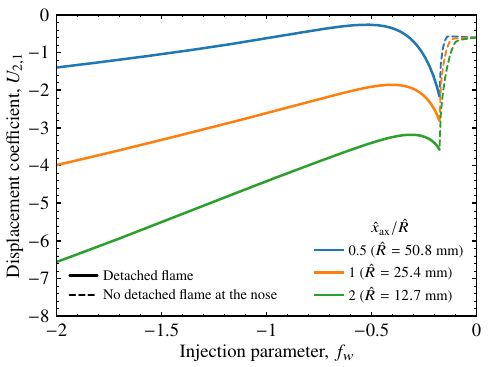}
    \caption{Displacement coefficient $\bm{U_{2,1}}$ of the variable-density model for uniform wall-normal injection over $\bm{\hat{x}_\mathrm{ax}/\hat{R} = 0.5}$, 1 and 2, and $\bm{Z_s = 0.228}$. Solid lines: flame sheet detached from the wall at the nose; dotted lines: $\bm{f_w > -0.1765}$, continuation of the cold layer without a detached sheet at the nose, outside the flame-sheet closure; vertical dotted line: onset of the detached flame.}
    \label{fig:U21}
\end{figure}

\Cref{fig:U21} shows the displacement coefficient. Without injection, the same value $U_{2,1} = \num{-0.601}$ is obtained for all porous extents. For weak injection, no detached flame exists at the nose, and the march gives $T = T_w$ wherever $Z_w < Z_s$, i.e., the cold layer. This part of the curve, shown dashed, continues outside the flame-sheet closure and does not represent a flame. Beyond a range of weak injection just after the flame detaches, $U_{2,1}$ becomes increasingly negative with injection, and the decrease is larger for a longer porous section (\cref{tab:flame}a). The heat release adds thermal displacement to the layer, and the effective strain is reduced more for a longer porous section. The thickening estimate $U_{2,1} \approx -\sqrt{U_{1,1}}\, \beta_1$ cannot reproduce this dependence on the porous extent. The Supplemental Material provides the march convergence and the contributions of the different body parts.

\subsection{Numerical Verification}\label{sec:cfd}
\begin{figure}[ht!]
    \centering
    \includegraphics{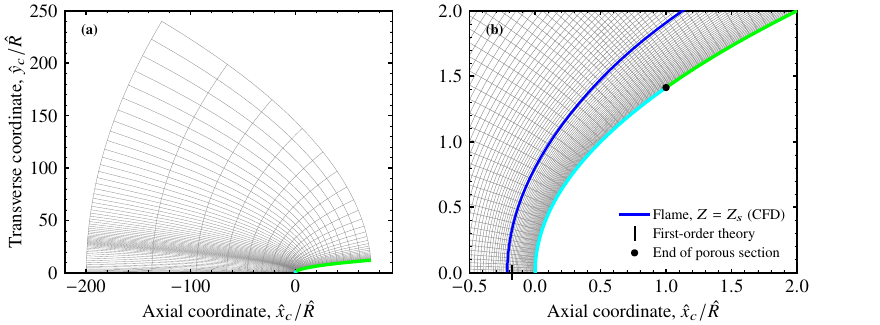}
    \caption{Coarse computational grid (\num{15360} cells) for $\bm{\Rey_R = 1000}$, $\bm{f_w = -1}$ and $\bm{\hat{x}_\mathrm{ax}/\hat{R} = 1}$, with the finer grids halving the spacing in both directions. (a)~Whole half-domain; (b)~nose region with all grid lines. The normal clustering scales with $\bm{\varepsilon}$. Blue line: flame sheet $\bm{Z = Z_s}$ of the numerical solution on the fine grid; black tick: first-order stand-off of the model; porous section in cyan and impermeable wall in green.}
    \label{fig:cfd_grid}
\end{figure}

The variable-density model is verified against two-dimensional steady laminar solutions of the same idealized problem. These solutions are computed with OpenFOAM \cite{weller_tensorial_1998} and a steady low-Mach solver developed for this study, with $\rho (1 + T) = 1$, $\rho \mu = 1$, a Burke-Schumann flame, the Robin condition for $Z$, and $\Pr = \Sc = \num{0.7}$. The orthogonal grids follow the parabolic coordinates of the body, and the cells normal to the wall are clustered with a spacing that scales with $\varepsilon$, with half of them within $30 \varepsilon$ of the wall (\cref{fig:cfd_grid}). Three nested grids are used, with \num{15360}, \num{61440} and \num{245760} cells, and each finer grid halves the spacing in both directions. The reported values are obtained by Richardson extrapolation from these grids. The inflow boundary is the line $\eta_p = 20$, about 200 nose radii upstream of the vertex. The potential flow displaced by the cold Blasius layer and a nose source equal to the injected flux are imposed there. No value of $U_{2,1}$ from the theory is used in the computations. The outlet is the line $\xi = 12$, which meets the body 72 nose radii downstream. The porous section is a fuel inlet with the Robin condition for $Z$, and the rest of the body is an impermeable wall with zero mixture-fraction flux. Both parts are held at temperature $T_w$. The Supplemental Material summarizes the boundary conditions. The computations cover $\Rey_R = 500$ to \num{2e4} for $f_w = -0.5$, $-1$ and $-2$ with $\hat{x}_\mathrm{ax}/\hat{R} = \num{0.5}$ and 1, and for $f_w = -1$ with $\hat{x}_\mathrm{ax}/\hat{R} = 2$. A fuel with \SI{20}{\percent} \ce{CH4} is also computed at $\Rey_R = 2000$ and 4000.

\begin{figure}[ht!]
    \centering
    \includegraphics{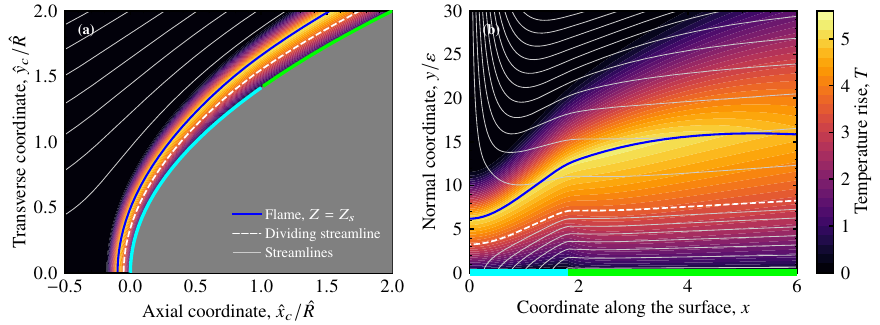}
    \caption{Temperature rise of the numerical solution on the fine grid for $\bm{\Rey_R = 4000}$, $\bm{f_w = -1}$ and $\bm{\hat{x}_\mathrm{ax}/\hat{R} = 1}$. (a)~Nose region in the physical plane; (b)~boundary layer against the distance along the surface, $\bm{x}$, and the distance from the wall along the lines of constant $\bm{\xi}$, $\bm{y}$, scaled with $\bm{\varepsilon}$. Thin lines: streamlines; dashed line: dividing streamline between the injected fluid and the free stream; blue line: flame sheet $\bm{Z = Z_s}$. Porous section in cyan and impermeable wall in green.}
    \label{fig:cfd_solution}
\end{figure}

A representative solution, for $\Rey_R = 4000$, $f_w = -1$ and $\hat{x}_\mathrm{ax}/\hat{R} = 1$, is shown in \cref{fig:cfd_solution}. The dividing streamline bounds the injected fluid, and the flame lies on its oxidizer side. The streamlines of the free stream cross the flame and turn along the body. The hot layer thickens along the porous section, and the flame remains detached downstream, where the layer no longer receives fuel. Near the stagnation line, the flame distance from the wall changes slowly along the surface.

The total numerical uncertainty, including the outer boundary, is below \SI{1}{\percent} for the stand-off distance and \SI{2.7}{\percent} for the wall shear and the scalar dissipation rate, except for the longer porous section at $\Rey_R \leqslant 1000$. The first-order errors decrease approximately as $\varepsilon$, whereas the two-term errors decrease approximately as $\varepsilon^2$, except for the scalar dissipation rate in the shorter porous section. For $\Rey_R \geqslant 1000$, the two-term errors in the stand-off distance, wall shear, scalar dissipation rate, and flame temperature are smaller than the first-order ones. Details on the error analysis, calibration of \cref{eq:validity}, and testing on held-out families are provided in the Supplemental Material. The correlation with $C = \num{1.3}$ underestimates the two-term errors of the shorter porous section for weak injection by up to a factor of two, and hence it is not a general bound. The solutions verify the laminar model. However, they do not show that the physical flow is laminar. A displacement thickness growing as in a turbulent layer downstream of the nose changes $U_{2,1}$ by an amount of the order of the second-order term (Supplemental Material).

\subsection{Flame Response to Curvature, Injection and Porous Extent}\label{sec:flame_response}
Each flame quantity is given by the value $q_1$ of the planar porous-wall flame, corrected for curvature and displacement as $q = q_1 [1 + \varepsilon (\kappa_{q,c} + U_{2,1} \kappa_{q,d})]$. The curvature part $\kappa_{q,c}$ and the displacement part $\kappa_{q,d}$ depend on $f_w$ and on the mixture, whereas the body enters only through $U_{2,1}$. \Cref{tab:flame} lists the first-order values and both parts for the reference mixture. Since $U_{2,1} < 0$ once the flame detaches, the displacement part has the same sign as the curvature part for every quantity in the table. Therefore, curvature and displacement act on the flame in the same direction.
\begin{table}[t!]
    \caption{\label{tab:flame} Flame response of the variable-density model for $\bm{Z_s = 0.228}$, $\bm{\gamma = 5.76}$, $\bm{T_w = 0}$ and $\bm{\Pr = 0.7}$. (a)~Displacement coefficient and first-order flame quantities; (b)~relative corrections, $\bm{q = q_1 [1 + \varepsilon (\kappa_{q,c} + U_{2,1} \kappa_{q,d})]}$, split into the curvature part $\bm{\kappa_{q,c}}$ and the displacement part $\bm{\kappa_{q,d}}$, for the stand-off distance, flame temperature, scalar dissipation rate, wall shear, wall heat flux and wall mixture fraction.}
    \centering
    \small
    \setlength{\tabcolsep}{4pt}
    \begin{tabular}{lrrrrrr}
        \toprule
        \multicolumn{7}{l}{(a)~Displacement coefficient and first-order quantities} \\
        \midrule
         & \multicolumn{3}{c}{$U_{2,1}$} & & & \\
        \cmidrule(lr){2-4}
        $f_w$ & $\hat{x}_\mathrm{ax}/\hat{R} = \num{0.5}$ & 1 & 2 & $y_{f1}/\varepsilon$ & $T_{f1}$ & $\hat{\chi}_{s1}/\hat{a}$ \\
        \midrule
        \num{-0.5} & \num{-0.26} & \num{-1.91} & \num{-3.40} & \num{2.85} & \num{3.68} & \num{0.139} \\
        \num{-1.0} & \num{-0.62} & \num{-2.61} & \num{-4.42} & \num{5.56} & \num{5.08} & \num{0.241} \\
        \num{-2.0} & \num{-1.40} & \num{-3.99} & \num{-6.56} & \num{8.64} & \num{5.69} & \num{0.307} \\
        \bottomrule
        \end{tabular}
        \par\medskip
        \begin{tabular}{llrrrrrr}
        \toprule
        \multicolumn{8}{l}{(b)~Curvature and displacement parts of the relative corrections} \\
        \midrule
        $f_w$ & Part & $y_f$ & $T_f$ & $\hat{\chi}_s$ & $\hat{\tau}_w$ & $\hat{q}_w$ & $Z_w$ \\
        \midrule
        \num{-0.5} & $\kappa_{q,c}$ & \num{3.65} & \num{0.76} & \num{-5.98} & \num{-5.73} & \num{-1.34} & \num{0.74} \\
         & $\kappa_{q,d}$ & \num{-1.14} & \num{-0.36} & \num{0.46} & \num{1.55} & \num{0.65} & \num{-0.36} \\
        \addlinespace
        \num{-1.0} & $\kappa_{q,c}$ & \num{5.41} & \num{0.48} & \num{-11.76} & \num{-8.94} & \num{-3.61} & \num{1.03} \\
         & $\kappa_{q,d}$ & \num{-0.87} & \num{-0.14} & \num{0.73} & \num{1.85} & \num{1.04} & \num{-0.30} \\
        \addlinespace
        \num{-2.0} & $\kappa_{q,c}$ & \num{7.20} & \num{0.16} & \num{-18.40} & \num{-13.17} & \num{-13.72} & \num{0.50} \\
         & $\kappa_{q,d}$ & \num{-0.77} & \num{-0.03} & \num{0.91} & \num{2.43} & \num{2.95} & \num{-0.11} \\
        \bottomrule
    \end{tabular}
\end{table}

When the injection increases, the planar flame moves away from the wall, and its stand-off distance in viscous lengths triples between $f_w = -0.5$ and $-2$ (\cref{tab:flame}a). The flame temperature rises because heat loss to the cold wall decreases, and the scalar dissipation rate at the flame also increases. At the wall, the shear stress and the heat flux decrease, and the mixture fraction approaches unity (\cref{tab:validation}).

\begin{figure}[ht!]
    \centering
    \includegraphics{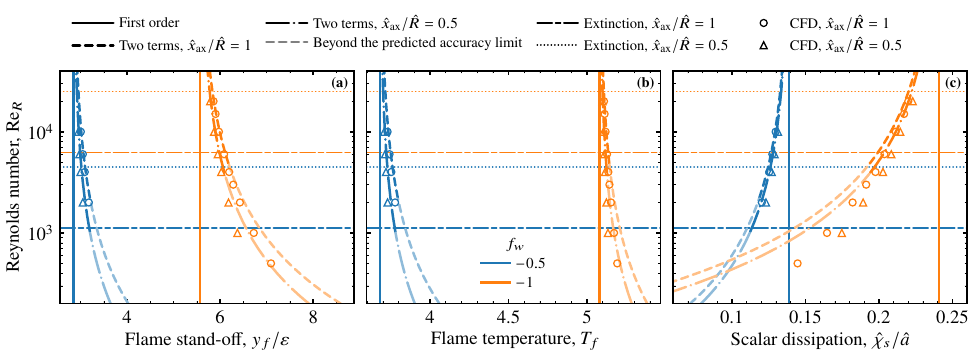}
    \caption{Flame quantities against the Reynolds number for variable density and $\bm{f_w = -0.5}$ and $\bm{-1}$. (a)~Stand-off distance in viscous lengths, $\bm{y_f/\varepsilon}$; (b)~flame temperature rise; (c)~stoichiometric scalar dissipation rate, $\bm{\hat{\chi}_s/\hat{a}}$. Lines: model, with the line styles of the legend, in light color beyond the accuracy limit predicted by \cref{eq:validity}; circles and triangles in the color of each $\bm{f_w}$: numerical solutions; horizontal lines: extinction of the finite-rate porous-wall flame in the color of each $\bm{f_w}$, for $\bm{\hat{R}} = \SI{25.4}{\milli\meter}$ and \SI{50.8}{\milli\meter} at $\bm{\hat{x}_\mathrm{ax}/\hat{R} = 1}$ and 0.5, respectively. The horizontal lines indicate planar extinction only, and they do not determine whether the cylinder flame burns.}
    \label{fig:flame_quantities}
\end{figure}

As the Reynolds number decreases, the boundary layer becomes thicker relative to the nose radius, and the flame moves away from the wall and becomes slightly hotter (\cref{fig:flame_quantities}). The scalar dissipation rate, wall shear, and wall heat flux decrease, whereas the wall mixture fraction increases. The relative changes in the scalar dissipation rate and wall shear are about twice those of the stand-off distance, and the change in flame temperature is about a tenth of that. E.g., for $f_w = -1$ and $\hat{x}_\mathrm{ax}/\hat{R} = 1$ at $\Rey_R = 4000$, the flame is located about \SI{12}{\percent} farther from the wall, and its scalar dissipation rate is about \SI{22}{\percent} lower than in the planar flame. Most of these corrections are due to curvature. For $\hat{x}_\mathrm{ax}/\hat{R} = 1$, displacement contributes about a third of the corrections to the stand-off distance and the wall shear, and about a sixth of the correction to the scalar dissipation rate. The curvature parts of the stand-off distance, the scalar dissipation rate, the wall shear, and the wall heat flux increase with injection (\cref{tab:flame}b). As a result, a flame farther from the wall is also more sensitive to nose curvature. However, the flame-temperature correction decreases with injection, since the flame approaches the limit without wall heat loss.

\begin{figure}[ht!]
    \centering
    \includegraphics{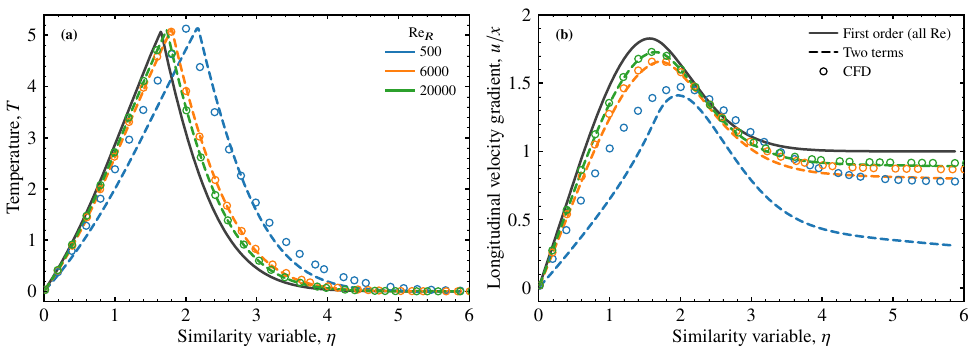}
    \caption{Stagnation-line profiles for variable density, $\bm{f_w = -1}$ and $\bm{\hat{x}_\mathrm{ax}/\hat{R} = 1}$. (a)~Temperature; (b)~longitudinal velocity gradient. Dark grey solid line: first-order model, same for all $\bm{\Rey_R}$; dashed lines: two-term expansion; symbols: numerical solutions.}
    \label{fig:cfd_profiles}
\end{figure}

The same response appears in the stagnation-line profiles of the flame structure in \cref{fig:cfd_profiles}. As $\Rey_R$ decreases, the flame moves away from the wall in the similarity variable, and its peak temperature changes little (\cref{fig:cfd_profiles}a). The velocity gradient above the layer falls below its first-order value, which corresponds to the reduction of the effective strain by displacement, and its peak inside the hot layer decreases (\cref{fig:cfd_profiles}b). The two-term profiles closely match the numerical solutions for $\Rey_R = 6000$ and \num{2e4}. At $\Rey_R = 500$, the corrections have the correct sign. However, they overpredict the flame shift.

As the porous section lengthens relative to the nose radius, $U_{2,1}$ becomes more negative, and the displacement parts of all corrections increase. For $f_w = -1$, when $\hat{x}_\mathrm{ax}/\hat{R}$ increases from \num{0.5} to 2, the correction of the wall shear increases by about \SI{70}{\percent} and that of the stand-off distance by about half, whereas the correction of the scalar dissipation rate increases by about a quarter. Therefore, the same porous section on a larger nose radius yields a flame closer to the planar porous-wall flame because of both the larger Reynolds number and the smaller displacement (\cref{fig:flame_quantities}).

When the fuel is diluted, $Z_s$ increases, and the flame moves toward the wall. For $f_w = -1$, the first-order stand-off distance for a fuel with \SI{15}{\percent} \ce{CH4} is about \SI{60}{\percent} of that of the reference mixture. The relative corrections of the stand-off distance change little with dilution at the same $f_w$ and $\Rey_R$ (Supplemental Material).

The flame sheet does not extinguish, and therefore the finite-rate porous-wall flame of \cref{sec:axi_models} provides a planar extinction reference. The extinction strain increases with injection, and the peak temperature decreases as the strain rate approaches extinction (Supplemental Material). From the reference mixture to the fuel with \SI{15}{\percent} \ce{CH4} at $f_w = -1$, the extinction strain decreases by more than an order of magnitude. Since extinction depends on the chemical time, the extinction Reynolds number, $\hat{a}_\mathrm{ext} \hat{R}^2/\hat{\nu}_\infty$, grows with the square of the nose radius. As a result, the planar extinction of the same porous section on a larger nose occurs at a higher $\Rey_R$, where the corrections are also smaller (\cref{fig:flame_quantities}). This comparison is a screening, not the cylinder's operating envelope, since planar extinction includes neither curvature nor displacement. For $\hat{x}_\mathrm{ax}/\hat{R} = \num{0.5}$, the computed two-term errors are within the tolerances below the planar-extinction Reynolds number for all sampled injection rates, although its boundary uncertainty is not measured. For $\hat{x}_\mathrm{ax}/\hat{R} = 1$, the overlap is narrow, and its ends depend on the constant of \cref{eq:validity} (Supplemental Material).

\subsection{Comparison with the Planar Slot Counterflow}\label{sec:slot}
\begin{figure}[t!]
    \centering
    \includegraphics{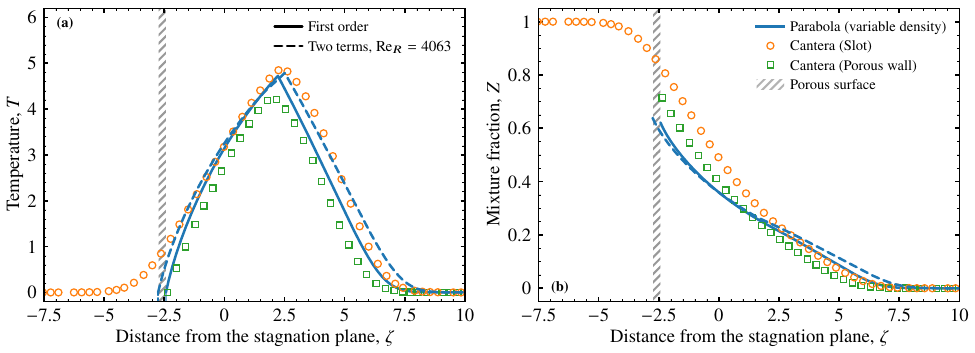}
    \caption{Parabola and slot flame of the reference mixture at $\bm{\hat{a}} = \SI{100}{\per\second}$ against the distance from the stagnation plane, $\bm{\zeta}$, in viscous lengths. (a)~Temperature rise; (b)~mixture fraction. Lines: parabola for $\bm{\Rey_R = 4063}$, i.e., $\bm{\hat{R}} = \SI{25.4}{\milli\meter}$, and $\bm{f_w = -0.8}$ (solid: first order; dashed: two terms), with the porous wall hatched; circles: detailed slot flame; squares: finite-rate porous-wall flame at the same $\bm{f_w}$.}
    \label{fig:slot}
\end{figure}

The parabola is compared with a detailed slot flame of the reference mixture at the same nominal strain rate in \cref{fig:slot}. This strain rate, $\hat{a} = \SI{100}{\per\second}$, sets the Damköhler number for the finite-rate flames and falls within the range of the experiments of Hu and Pitz \cite{hu_structural_2009a}. This comparison gives context for the effect of the porous wall, but it does not isolate curvature. The slot has a gap of \SI{11.8}{\milli\meter}, and both inlet velocities are scaled until the strain rate matches the nominal value (Supplemental Material). The parabola is computed for $\Rey_R = 4063$, i.e., $\hat{R} = \SI{25.4}{\milli\meter}$, with $\hat{x}_\mathrm{ax}/\hat{R} = 1$ and $f_w = -0.8$, and displacement reduces its effective strain rate by about \SI{4}{\percent}. The finite-rate porous-wall flame at the same $f_w$ and strain rate is also shown.

The profiles of \cref{fig:slot} are plotted against the distance $\zeta$ from the stagnation plane, in viscous lengths $(\hat{\nu}_\infty/\hat{a})^{1/2}$. The first-order parabola is the planar porous-wall flame sheet at the same mass flux and wall temperature. Thus, the difference between its first-order and two-term profiles isolates curvature and displacement, which move the flame away from the stagnation plane by \SI{13}{\percent}. The slot flame lies slightly farther from the stagnation plane, and its peak temperature rise exceeds that of the flame sheet by about \SI{2}{\percent}. Because the slot and porous-wall flames share the same detailed formulation, their difference mainly reflects the boundary conditions. The slot flame is hotter than the porous-wall flame, and its temperature rises upstream of the porous surface, where diffusion heats the fuel stream, whereas the porous wall is held at $T_w = 0$. The finite-rate porous-wall flame has the same planar geometry, injection, and wall as the first-order parabola, and it differs only in its physical formulation. Consequently, the difference between the squares and the first-order line in \cref{fig:slot} reflects only the formulation: detailed thermochemistry, variable specific heat and molecular weight, mixture-averaged transport with non-unity Lewis numbers, and finite reaction rate. The finite-rate temperature profile lies closer to the wall across the whole layer, and its surface $Z = Z_s$ lies about \SI{9}{\percent} closer to the wall, with a lower temperature rise (\cref{fig:slot}a). Its mixture fraction is higher at the wall, and it falls to zero closer to the stagnation plane (\cref{fig:slot}b). These differences increase with injection at the same strain rate (Supplemental Material). Therefore, at this strain rate, curvature and displacement move the flame by an amount comparable to the difference between the flame sheet and the detailed formulation, and in the opposite direction.

\section{Conclusion}
This work presents a second-order matched asymptotic description of a Burke-Schumann flame detached from the wall at the forward stagnation line of a porous parabolic cylinder. The model includes curvature, displacement, a Robin condition for the mixture fraction and thermal expansion, for unity Lewis numbers, $\Pr = \num{0.7}$, and a diluted methane-air mixture with $Z_S = \num{0.228}$. The nose radius sets the curvature at the stagnation line, whereas the rest of the body enters only through the displacement coefficient.

Results show that curvature and displacement act on the flame in the same direction. As the Reynolds number decreases, the flame moves away from the wall, and the scalar dissipation rate, wall shear, and wall heat flux decrease, whereas the flame temperature changes little. Most of these corrections stem from curvature. Stronger injection moves the flame away from the wall and increases its sensitivity to curvature. In addition, thermal expansion increases the wall shear curvature coefficient several-fold. The displacement coefficient depends on the injection along the whole porous section. As a result, a longer porous section reduces the effective strain more, and the same porous section on a larger nose gives a flame closer to the planar porous-wall flame. Fuel dilution brings the flame toward the wall and lowers its extinction strain considerably.

In the cases studied, the two-term errors against Navier-Stokes solutions of the same idealized problem decrease approximately as the square of the expansion parameter, except for the scalar dissipation rate of the shorter porous section, and the second-order terms considerably reduce the first-order theory errors at the Reynolds numbers of laboratory-scale burners. Screening against planar finite-rate extinction shows a narrow overlap with the accuracy limits for a nose radius equal to the axial length of the porous section, and an overlap across all sampled injection rates for a nose radius twice as large. In selected planar cases, the detailed finite-rate and transport formulation places the flame closer to the wall than the flame sheet, with a lower flame temperature, and this difference is comparable to the effect of curvature and displacement. However, the total error relative to experimental flames is not established, and the trends outside the tested families are extrapolations of these results.

\appendix
\section{Displacement Coefficient}\label[appendix]{app:U21}
In the parabolic coordinates, the fluid occupies $\eta_p > 1$. The first-order potential is $w_1 = (\zeta - \mathrm{i})^2/2$, with $\zeta = \xi + \mathrm{i} \eta_p$, and its stream function, $\psi_1 = \xi (\eta_p - 1)$, increases away from the body. The second-order stream function $\psi_2$ is harmonic in $\eta_p > 1$, and the boundary layer displaces the streamlines by the volume flux $Q$,
\begin{equation}\label{eq:app_psi2}%
    \psi_2(\xi, 1) = - Q(\xi)
    \qquad\text{and}\qquad
    Q = U_e \delta^* - \psi_w = \xi \Delta - \psi_w
\end{equation}
where $\Delta$ excludes the injected fluid, which enters through $\psi_w$. The flux $Q$ is odd in $\xi$. Its part $\xi \Delta_\infty$ gives $\psi_2 = -\Delta_\infty \xi$, i.e., the confocal parabola $\eta_p = 1 + \varepsilon \Delta_\infty$, without slip velocity. The derivative of the remainder, $\tilde{Q} = \xi \tilde{\Delta} - \psi_w$, vanishes downstream. For a function analytic in the half plane, $\partial_\xi \phi_2 = -\mathcal{H}[\partial_\xi \psi_2]$ on the boundary, where $\phi_2$ is the potential and $\mathcal{H}[q](\xi) = \pi^{-1}\, \mathrm{p.v.} \int_{-\infty}^{\infty} q(t)/(\xi - t)\, \mathrm{d}t$. The slip velocity is $U_2 = (1 + \xi^2)^{-1/2}\, \mathcal{H}[\tilde{Q}']$, with primes denoting $\mathrm{d}/\mathrm{d}\xi$ in this appendix, and hence
\begin{equation}\label{eq:app_U21}%
    U_{2,1} = \frac{\mathrm{d} U_2}{\mathrm{d} \xi}\bigg|_{\xi = 0} =
    - \frac{1}{\pi} \int_{-\infty}^{\infty} \frac{\tilde{Q}''(t)}{t}\, \mathrm{d}t =
    - \frac{2}{\pi} \int_0^\infty \frac{\tilde{Q}''(t)}{t}\, \mathrm{d}t
\end{equation}
The integrand is regular at the origin, since $\tilde{\Delta}$ is even. With $\int_0^\infty (t \tilde{\Delta})''/t\, \mathrm{d}t = 2 \int_0^\infty \tilde{\Delta}'/t\, \mathrm{d}t$, the layer gives the first term of \cref{eq:U21}. The second derivative of $\psi_w = f_w \min(x, x_p)$ is $f_w \xi/(1 + \xi^2)^{1/2}$ for $\xi < \xi_p$, plus $-f_w (1 + \xi_p^2)^{1/2} \delta_D(\xi - \xi_p)$ at the edge of the porous section, where $\delta_D$ is the Dirac delta. These give the terms $\operatorname{arcsinh} \xi_p$ and $-(1 + \xi_p^2)^{1/2}/\xi_p$ of \cref{eq:U21}. The march computes $q = \Delta - F_w$, with $F_w = \psi_w/\xi$, and $Q = \xi q$. The same steps give $U_{2,1} = -(4/\pi) \int_0^\infty q'(t)/t\, \mathrm{d}t$, which equals \cref{eq:U21}. A change $\delta q$ that vanishes for $\xi < \xi_t$ changes the coefficient by
\begin{equation}\label{eq:app_dU21}%
    \delta U_{2,1} =
    - \frac{4}{\pi} \int_{\xi_t}^\infty \frac{\delta q'(t)}{t}\, \mathrm{d}t =
    - \frac{4}{\pi} \int_{\xi_t}^\infty \frac{\delta q(t)}{t^2}\, \mathrm{d}t
\end{equation}
i.e., the displacement at $\xi$ has the weight $4/(\pi \xi^2)$.

\section{Inner Problem}\label[appendix]{app:inner}
\textbf{Flame-sheet conditions.} The first-order problem is solved on $0 \leqslant \eta \leqslant \eta_{f1}$ and $\eta_{f1} \leqslant \eta \leqslant 16$, with $g(\eta_{f1}) = Z_s$. The functions $f$, $f'$, $f''$, $g$, $g'$, $h$, $h'$ and $N_0$ are continuous at $\eta_{f1}$, and so are $\theta$ and $f'''$. The second-order functions $F_k$, $F'_k$, $F''_k$, $G_k$, $G'_k$, $H_k$, $H'_k$ and $N_1$, with $k = c, d$, are also continuous. The temperature $\theta_k$ uses the one-sided value of $\partial T/\partial Z$. With $[q] = q(\eta_{f1}^+) - q(\eta_{f1}^-)$ and $\Gamma = \gamma/[Z_s (1 - Z_s)]$, the jumps are $[\theta_k] = \Gamma G_k$ and $[\theta'] = \Gamma g'$, and hence \cref{eq:Fd,eq:Fc} give $[F'''_d] = -\Gamma G_d$ and $[F'''_c] = \Gamma (f' g' - G_c)$, all evaluated at $\eta_{f1}$. The expansion of $T$ at fixed $\eta$ is not uniformly valid within $\mathcal{O}(\varepsilon)$ of the sheet, which changes integrals of $\theta$ only by $\mathcal{O}(\varepsilon^2)$. The sheet shift enters through $\eta_{f2}$ in \cref{eq:yf2,eq:Tf2,eq:chi2}, which contain only continuous functions.

\textbf{Scalar dissipation rate.} With $\rho = 1/\theta$, $\mu = \theta$ and unity Lewis number, $\hat{\mathcal{D}} = \hat{\nu}_\infty \theta^2/\Pr$. At the stagnation line, the tangential derivative of $Z$ vanishes, and $\partial Z/\partial \hat{n} = (\hat{a}/\hat{\nu}_\infty)^{1/2}\, \partial_\eta Z/\theta$ for $U_{1,1} = 1$. Hence,
\begin{equation}\label{eq:app_chi}%
    \hat{\chi} =
    2 \hat{\mathcal{D}} \left( \frac{\partial Z}{\partial \hat{n}} \right)^2 =
    2\, \frac{\hat{\nu}_\infty \theta^2}{\Pr}\, \frac{\hat{a}}{\hat{\nu}_\infty}\, \frac{(\partial_\eta Z)^2}{\theta^2} =
    \frac{2 \hat{a}}{\Pr} (\partial_\eta Z)^2
\end{equation}

\textbf{Wall shear.} At the wall, $u = 0$ and $\partial v/\partial x = 0$, and $\mu_w\, \partial \eta/\partial y = \sqrt{U_{1,1}}/\varepsilon$. For $U_{1,1} = 1$,
\begin{equation}\label{eq:app_tau}%
    \frac{\hat{\tau}_w}{\hat{\rho}_\infty \hat{U}_\infty^2} =
    \varepsilon x \left[ f''(0) + \varepsilon \left( F''_c(0) + U_{2,1} F''_d(0) \right) \right] + \mathcal{O}(\varepsilon^3 x, \varepsilon x^3)
\end{equation}
The verification uses the coefficient of $x$, obtained from the numerical solutions by fitting $u/x$ on the stagnation line with a polynomial over $0 < \eta < 1$.

\section*{Funding Sources}
This work was supported by the National Science Foundation through grant \#2302003.

\section*{Acknowledgments}
The authors thank F.F. Fachini for fruitful discussions. The authors acknowledge using Claude (Anthropic), a large language model assistant, to proofread the derivation and the manuscript and to improve the code. All AI-assisted text and code were reviewed, tested, and verified by the authors, who take full responsibility for the content of this paper.


\clearpage
\begin{center}{\Large\bfseries Supplemental Material}\end{center}
\setcounter{section}{0}\setcounter{figure}{0}\setcounter{table}{0}\setcounter{equation}{0}
\renewcommand{\thesection}{S\arabic{section}}
\renewcommand{\thesubsection}{\thesection.\arabic{subsection}}
\renewcommand{\thefigure}{S\arabic{figure}}
\renewcommand{\thetable}{S\arabic{table}}
\renewcommand{\theequation}{S\arabic{equation}}

This Supplemental Material provides the numerical details used to verify the manuscript and to assess the accuracy of the two-term expansion. \Cref{sec:s_slot} verifies the slot counterflow solver. \Cref{sec:s_uncertainty} describes the numerical uncertainty of the Navier-Stokes solutions and the tests of their outer boundary. \Cref{sec:s_accuracy} discusses model errors relative to these solutions, error correlation, and accuracy limits relative to extinction of the finite-rate porous-wall flame. \Cref{sec:s_march} describes the reacting march and its convergence, \cref{sec:s_porous} the extinction estimates of the finite-rate porous-wall flame, and \cref{sec:s_repro} the codes and data. Equation numbers without the prefix S refer to the manuscript, and the notation is that of the manuscript.

\section{Verification of the Slot Counterflow Solver}\label[section]{sec:s_slot}
The slot solutions use a modified Cantera code \cite{goodwin_cantera_2024}, in which the factor of the continuity equation of the axisymmetric opposed flow is changed for planar flow \cite{kee_chemically_2003}. The code is based on Cantera 3.1.0, and it is available in the branch \texttt{tubular} of the repository \texttt{https://github.com/vsauer/cantera}.

\begin{figure}[ht!]
    \centering
    \includegraphics{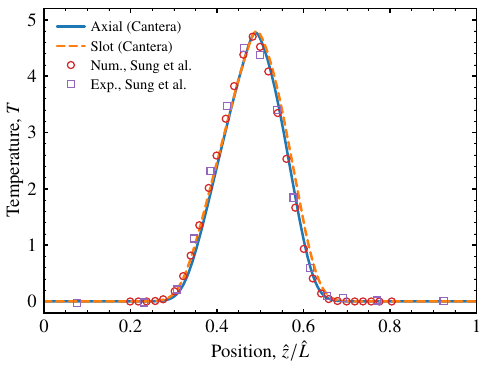}
    \caption{Temperature for Case B of Sung et al.~\cite{sung_structural_1995} against the distance $\bm{\hat{z}}$ from the fuel inlet, scaled with the gap $\bm{\hat{L}}$. Lines: Cantera axial and slot solutions; symbols: computations and measurements of \cite{sung_structural_1995}.}
    \label{fig:s_sung}
\end{figure}

\textbf{Case B of Sung et al.} The axial and slot solutions are compared with Case B of Sung et al.~\cite{sung_structural_1995}, a flame of \SI{23}{\percent} \ce{CH4} in \ce{N2} against \SI{23}{\percent} \ce{O2} in \ce{N2} by volume. The nozzles are \SI{13}{\milli\meter} apart, and the inlet velocities give a strain rate close to that of Sung et al., who chose the inlet mass fluxes to align computed and measured temperatures. The solutions use the San Diego mechanism \cite{ucsd_san_2016} and mixture-averaged transport, and they are insensitive to further grid refinement. The temperature profiles in \cref{fig:s_sung} nearly coincide under the same inlet conditions, and they agree with the computations and measurements of Sung et al. The normal velocity profiles also nearly coincide. However, the strain rate from the eigenvalue, $\hat{a} = (-\hat{\Lambda}/\hat{\rho}_\infty)^{1/2}$, is close to the value of Sung et al.\ for the axial flame and about 1.6 times larger for the slot flame. This eigenvalue measures the tangential velocity gradient, which equals the normal strain in planar flow and half of it in axisymmetric flow. Therefore, the two values differ even though the flames nearly coincide.

\textbf{Planar benchmark.} For isothermal, non-reacting nitrogen streams with the gap and the inlet velocities of the slot flame of the manuscript, the density and viscosity are uniform. The finite-gap equations of the manuscript with $b = 0$ then reduce to
\begin{equation}\label{eq:s_bvp}
    \frac{\mathrm{d} \hat{u}_z}{\mathrm{d} \hat{z}} = - \hat{W}
    \qquad\text{and}\qquad
    \hat{\rho} \left( \hat{u}_z \frac{\mathrm{d}\hat{W}}{\mathrm{d}\hat{z}} + \hat{W}^2 \right) =
    - \hat{\Lambda} + \hat{\mu} \frac{\mathrm{d}^2 \hat{W}}{\mathrm{d}\hat{z}^2}
\end{equation}
with the inlet velocities and $\hat{W} = 0$ at both inlets, and $\hat{\Lambda}$ as an unknown parameter. This problem is solved separately with the collocation solver of SciPy \cite{kierzenka_bvp_2001,virtanen_scipy_2020}. As the uniform Cantera grid is refined, its eigenvalue converges to that of the boundary-value problem at first order, and the Richardson extrapolation agrees with it to round-off level. On the finest grid, the velocity profiles agree to a small fraction of a percent, and the axisymmetric solution ($b = 1$) agrees to the same level. For the reacting slot flame of the manuscript, the lateral outflow balances the inlet mass fluxes, and the outflows of C, H, O and N equal their inflows within \SI{0.2}{\percent}.

\section{Numerical Uncertainty of the Navier-Stokes Solutions}
\begin{figure}[ht!]
    \centering
    \includegraphics{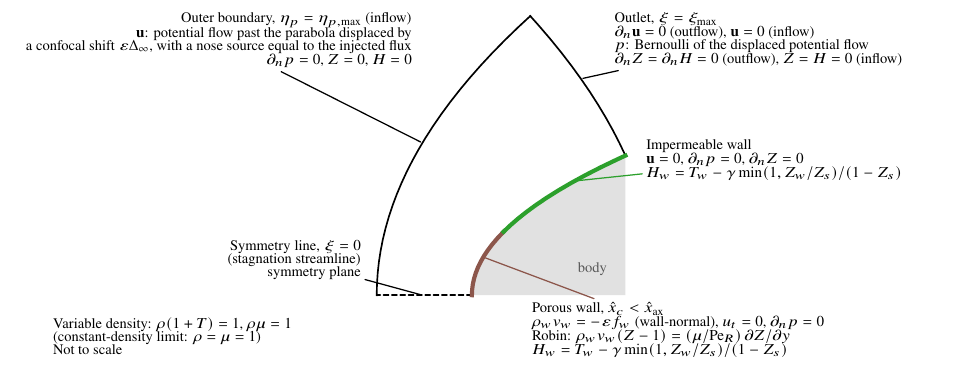}
    \caption{Boundary conditions of the numerical solutions, with the domain not to scale. Porous section in brown and impermeable wall in green. Here, $\bm{\mathbf{u}}$ is the velocity vector, $\bm{u_t}$ its component tangential to the wall, $\bm{\partial_n}$ the derivative normal to the boundary, $\bm{Z_w}$ the wall value of the mixture fraction, and $\bm{\eta_{p,\max}}$ and $\bm{\xi_{\max}}$ the parabolic coordinates of the inflow boundary and the outlet.}
    \label{fig:s_bcs}
\end{figure}

\label[section]{sec:s_uncertainty}
\textbf{Grids and boundary conditions.} The grids are orthogonal and follow the parabolic coordinates of the body, as shown in the manuscript together with a representative solution. The boundary conditions are summarized in \cref{fig:s_bcs}.

\textbf{Grid uncertainty.} With the values $q_2$, $q_3$ and $q_4$ of a quantity on the three grids, $d_1 = q_3 - q_2$ and $d_2 = q_4 - q_3$, one rule is used for every case. For monotone convergence, $d_1 d_2 > 0$ and $|d_2| < |d_1|$, with an observed order $p = \log_2 |d_1/d_2|$ between 1 and 4, the uncertainty is the fine-grid convergence index, $1.25\, |d_2|/(2^p - 1)$ \cite{roache_perspective_1994}. For any other sequence, i.e., oscillatory, not converging, or with $p$ outside $[1, 4]$, the uncertainty is $\max(|d_1|, |d_2|)$, which corresponds to the index with the safety factor 3 and the assumed order 2 applied to the larger change. The reported values retain the Richardson extrapolation of the post-processing, with the observed order clipped to $1 \leqslant p \leqslant 4$ for monotone sequences and the assumed order 2 for the others. The conservative uncertainty above is computed separately. The grid uncertainty of the flame quantities is small, and it exceeds \SI{1}{\percent} only for the scalar dissipation rate at a few strongly injected cases.

\textbf{Outer boundary and outlet.} The inflow boundary is the line $\eta_p = 20$, about 200 nose radii upstream of the vertex, and the outlet is $\xi = 12$, which meets the body 72 nose radii downstream. The inflow condition uses the cold Blasius displacement, whereas the hot layer is thicker, which gives a confinement error of order $\varepsilon$. This error is measured at $\Rey_R = \num{e4}$ and $f_w = -1$ by moving the inflow boundary and the outlet further away, on grids that keep the original cells. The change induced by the inflow boundary decays approximately as $\eta_p^{-3}$. For $\hat{x}_\mathrm{ax}/\hat{R} = 1$, the remaining inflow and outlet errors are small fractions of the two-term errors, and they have opposite signs. For $\hat{x}_\mathrm{ax}/\hat{R} = 2$, the outlet dominates, and its effect on the stand-off distance reaches about half of the two-term error. The outer-boundary uncertainty is taken as twice the change from $\eta_p = 20$ to 40 plus twice the change from $\xi = 12$ to 16, which bounds both measured errors. At other values of $\Rey_R$ and $f_w$, it is scaled with the change produced by moving the inflow boundary from $\eta_p = 20$ to 40. No boundary tests were run for $\hat{x}_\mathrm{ax}/\hat{R} = 0.5$, and its uncertainty is taken from the family with $\hat{x}_\mathrm{ax}/\hat{R} = 1$ and the same $f_w$. This transferred value is an allowance, not a measured bound. Hence, the accuracy results of the shorter porous section are provisional. This adds to the grid uncertainty, giving the total uncertainty $N_u$. A value enters the two-term error exponents and the calibration of $C$ only if $N_u \leqslant |e_2|/3$, where $e_2$ is the two-term error relative to the computed value. The first-order fits use all values in the calibration domain. The fits weight $\log |e_i|$ with $\sigma_i^{-2}$, where $\sigma_i = \max(N_{u,i}/|e_i|, 0.01)$, and assume uncorrelated uncertainties, although the domain allowance can be correlated between Reynolds numbers. Hence, their $\pm$ values are formal standard errors under this weighting, and they are not confidence limits of the asymptotic exponent.

\textbf{Balances and residuals.} The global mass and mixture-fraction balances close to round-off level on the finest grids, and all computations reach initial residuals below \num{e-11}. The mixture-fraction flux through the inflow boundary is negligible, and the outlet carries the injected mixture fraction.

\Cref{tab:gci} lists the observed orders, the uncertainties, and their ratio to the two-term error. \Cref{tab:slopes} lists the error exponents. For the longer injection length, the stand-off distance is unresolved at $\Rey_R \geqslant 2000$.
\begin{table}[ht!]
    \caption{\label{tab:gci} Numerical uncertainty of the variable-density computations. For each quantity: observed order $\bm{p}$ of the three grids (osc., oscillatory; n.c., not converging; parentheses, outside $\bm{1 \leqslant p \leqslant 4}$), total uncertainty $\bm{N_u}$ (grid and outer boundary) in percent of the computed value, and its ratio to the two-term error $\bm{|e_2|}$. The extrapolated values follow the post-processing policy: the observed order is clipped to $\bm{[1, 4]}$ for monotone sequences and set to 2 for the others, whereas the conservative $\bm{N_u}$ is computed separately. Values with $\bm{N_u > |e_2|/3}$, marked with an asterisk, are not used for the two-term error slopes and for $\bm{C}$. The computations with $\bm{\hat{x}_\mathrm{ax}/\hat{R} = 2}$ at $\bm{\Rey_R \leqslant 1000}$ lie outside the calibration domain.}
    \centering
    \footnotesize
    \setlength{\tabcolsep}{3.2pt}
    \begin{tabular}{lr rrr rrr rrr}
        \toprule
         & & \multicolumn{3}{c}{$y_f$} & \multicolumn{3}{c}{$\tau_w$} & \multicolumn{3}{c}{$\chi_s$} \\
        \cmidrule(lr){3-5} \cmidrule(lr){6-8} \cmidrule(lr){9-11}
        Family & $\Rey_R$ & $p$ & $N_u$ (\%) & $N_u/|e_2|$ & $p$ & $N_u$ (\%) & $N_u/|e_2|$ & $p$ & $N_u$ (\%) & $N_u/|e_2|$ \\
        \midrule
        $f_w = -0.5$ & \num{2000} & \num{1.81} & \num{0.126} & \num{0.09} & \num{1.97} & \num{1.032} & \num{0.19} & n.c. & \num{0.298} & \num{0.13} \\
         & \num{4000} & \num{1.87} & \num{0.087} & \num{0.13} & \num{1.97} & \num{0.998} & \num{0.36}$^*$ & \num{1.49} & \num{0.109} & \num{0.09} \\
         & \num{6000} & \num{1.92} & \num{0.070} & \num{0.15} & \num{2.04} & \num{0.891} & \num{0.47}$^*$ & n.c. & \num{0.620} & \num{0.86}$^*$ \\
         & \num{10000} & \num{1.81} & \num{0.062} & \num{0.22} & \num{2.04} & \num{0.877} & \num{0.76}$^*$ & \num{1.03} & \num{0.528} & \num{1.79}$^*$ \\
        \addlinespace
        $f_w = -1$ & \num{500} & \num{2.22} & \num{0.678} & \num{0.12} & \num{1.89} & \num{1.987} & \num{0.04} & n.c. & \num{1.350} & \num{0.04} \\
         & \num{1000} & \num{1.76} & \num{0.467} & \num{0.16} & \num{1.96} & \num{1.459} & \num{0.06} & \num{1.18} & \num{0.733} & \num{0.04} \\
         & \num{2000} & \num{1.42} & \num{0.329} & \num{0.22} & \num{1.85} & \num{1.223} & \num{0.10} & \num{1.12} & \num{0.729} & \num{0.09} \\
         & \num{3000} & \num{1.74} & \num{0.257} & \num{0.25} & \num{1.95} & \num{1.015} & \num{0.12} & \num{2.94} & \num{0.218} & \num{0.04} \\
         & \num{4000} & \num{1.99} & \num{0.216} & \num{0.28} & \num{1.95} & \num{0.945} & \num{0.15} & \num{2.41} & \num{0.207} & \num{0.05} \\
         & \num{6000} & \num{1.75} & \num{0.179} & \num{0.35}$^*$ & \num{1.92} & \num{0.881} & \num{0.21} & n.c. & \num{0.789} & \num{0.30} \\
         & \num{10000} & \num{1.83} & \num{0.133} & \num{0.42}$^*$ & \num{1.94} & \num{0.791} & \num{0.32} & \num{1.22} & \num{0.301} & \num{0.19} \\
         & \num{15000} & \num{1.89} & \num{0.106} & \num{0.50}$^*$ & \num{1.93} & \num{0.749} & \num{0.46}$^*$ & \num{1.68} & \num{0.161} & \num{0.12} \\
         & \num{20000} & \num{2.04} & \num{0.089} & \num{0.54}$^*$ & \num{1.92} & \num{0.723} & \num{0.59}$^*$ & n.c. & \num{0.590} & \num{0.57}$^*$ \\
        \addlinespace
        $f_w = -2$ & \num{3000} & \num{1.35} & \num{0.723} & \num{0.39}$^*$ & (\num{4.06}) & \num{2.202} & \num{0.10} & \num{1.22} & \num{1.092} & \num{0.07} \\
         & \num{6000} & osc. & \num{0.510} & \num{0.52}$^*$ & \num{2.52} & \num{1.517} & \num{0.14} & n.c. & \num{1.236} & \num{0.17} \\
         & \num{10000} & osc. & \num{0.384} & \num{0.63}$^*$ & \num{1.45} & \num{1.189} & \num{0.18} & \num{3.00} & \num{0.479} & \num{0.11} \\
         & \num{20000} & osc. & \num{0.261} & \num{0.82}$^*$ & \num{1.97} & \num{0.825} & \num{0.24} & n.c. & \num{1.727} & \num{0.86}$^*$ \\
        \addlinespace
        $\hat{x}_\mathrm{ax}/\hat{R} = 2$, $f_w = -1$ & \num{500} & \num{1.98} & \num{1.718} & \num{0.45} & \num{1.83} & \num{4.111} & \num{0.07} & \num{2.05} & \num{2.890} & \num{0.07} \\
         & \num{1000} & \num{1.97} & \num{1.175} & \num{0.60} & \num{1.91} & \num{2.959} & \num{0.09} & n.c. & \num{2.642} & \num{0.14} \\
         & \num{2000} & \num{2.47} & \num{0.802} & \num{0.81}$^*$ & \num{1.97} & \num{2.174} & \num{0.14} & n.c. & \num{2.289} & \num{0.26} \\
         & \num{4000} & \num{2.15} & \num{0.550} & \num{1.76}$^*$ & \num{1.86} & \num{1.713} & \num{0.23} & n.c. & \num{1.461} & \num{0.35}$^*$ \\
         & \num{10000} & \num{2.75} & \num{0.332} & \num{3.50}$^*$ & \num{1.93} & \num{1.223} & \num{0.42}$^*$ & (\num{0.31}) & \num{1.253} & \num{0.95}$^*$ \\
         & \num{20000} & \num{2.09} & \num{0.229} & \num{7.27}$^*$ & \num{1.92} & \num{1.025} & \num{0.75}$^*$ & \num{1.25} & \num{0.555} & \num{0.65}$^*$ \\
        \addlinespace
        \SI{20}{\percent} \ce{CH4}, $f_w = -1$ & \num{2000} & \num{1.83} & \num{0.321} & \num{0.20} & \num{1.86} & \num{1.219} & \num{0.13} & \num{1.47} & \num{0.309} & \num{0.06} \\
         & \num{4000} & \num{1.60} & \num{0.225} & \num{0.28} & \num{1.95} & \num{0.944} & \num{0.19} & \num{4.00} & \num{0.155} & \num{0.06} \\
        \addlinespace
        $\hat{x}_\mathrm{ax}/\hat{R} = 0.5$, $f_w = -0.5$ & \num{2000} & \num{1.77} & \num{0.134} & \num{0.08} & \num{1.97} & \num{1.068} & \num{0.24} & (\num{4.17}) & \num{1.355} & \num{0.52}$^*$ \\
         & \num{4000} & \num{1.85} & \num{0.094} & \num{0.10} & \num{2.04} & \num{0.922} & \num{0.39}$^*$ & \num{3.38} & \num{0.024} & \num{0.02} \\
         & \num{6000} & \num{1.89} & \num{0.078} & \num{0.13} & \num{2.04} & \num{0.904} & \num{0.57}$^*$ & \num{2.89} & \num{0.042} & \num{0.05} \\
         & \num{10000} & \num{1.93} & \num{0.063} & \num{0.17} & \num{2.04} & \num{0.899} & \num{0.95}$^*$ & \num{2.86} & \num{0.036} & \num{0.05} \\
        \addlinespace
        $\hat{x}_\mathrm{ax}/\hat{R} = 0.5$, $f_w = -1$ & \num{1000} & \num{1.71} & \num{0.477} & \num{0.13} & \num{1.83} & \num{1.616} & \num{0.09} & \num{2.14} & \num{0.334} & \num{0.02} \\
         & \num{2000} & \num{1.82} & \num{0.324} & \num{0.16} & \num{1.92} & \num{1.208} & \num{0.13} & \num{1.25} & \num{0.431} & \num{0.06} \\
         & \num{4000} & \num{1.86} & \num{0.222} & \num{0.21} & \num{1.93} & \num{0.989} & \num{0.20} & \num{1.74} & \num{0.232} & \num{0.06} \\
         & \num{6000} & \num{1.95} & \num{0.177} & \num{0.24} & \num{1.92} & \num{0.908} & \num{0.27} & n.c. & \num{0.723} & \num{0.28} \\
         & \num{10000} & \num{1.76} & \num{0.138} & \num{0.31} & \num{1.92} & \num{0.826} & \num{0.41}$^*$ & n.c. & \num{0.987} & \num{0.71}$^*$ \\
         & \num{20000} & \num{1.73} & \num{0.098} & \num{0.43}$^*$ & \num{2.02} & \num{0.670} & \num{0.64}$^*$ & \num{3.90} & \num{0.075} & \num{0.06} \\
        \addlinespace
        $\hat{x}_\mathrm{ax}/\hat{R} = 0.5$, $f_w = -2$ & \num{2000} & \num{2.01} & \num{0.908} & \num{0.28} & \num{1.76} & \num{2.698} & \num{0.12} & \num{3.09} & \num{0.501} & \num{0.03} \\
         & \num{4000} & \num{2.60} & \num{0.621} & \num{0.35}$^*$ & \num{1.24} & \num{1.873} & \num{0.16} & n.c. & \num{1.482} & \num{0.15} \\
         & \num{10000} & \num{1.47} & \num{0.384} & \num{0.50}$^*$ & \num{1.87} & \num{1.162} & \num{0.23} & \num{1.92} & \num{0.504} & \num{0.11} \\
         & \num{20000} & \num{1.89} & \num{0.261} & \num{0.64}$^*$ & \num{1.56} & \num{0.829} & \num{0.31} & \num{1.21} & \num{0.734} & \num{0.38}$^*$ \\
        \bottomrule
    \end{tabular}
\end{table}

\begin{table}[ht!]
    \caption{\label{tab:slopes} Exponent $\bm{n}$ of the relative errors, $\bm{|e| \propto \varepsilon^n}$, from least-squares fits weighted with $\bm{N_u}$, with the number of computations in parentheses. The $\bm{\pm}$ values are formal standard errors of the weighted fits under this weighting, and they are not confidence limits of the asymptotic exponent. Two-term fits use only the values with $\bm{N_u \leqslant |e_2|/3}$ (\cref{tab:gci}), and first-order fits use all values in the calibration domain. For two points, no fit error is given.}
    \centering
    \small
    \setlength{\tabcolsep}{4pt}
    \begin{tabular}{ll llll}
        \toprule
        Family & Range & $y_f$ & $\tau_w$ & $\chi_s$ & $T_f$ \\
        \midrule
        \multicolumn{6}{l}{\textit{First order}} \\
        $f_w = -0.5$ & all & \num{0.85} $\pm$ \num{0.02} (4) & \num{0.91} $\pm$ \num{0.13} (4) & \num{1.01} $\pm$ \num{0.06} (4) & \num{0.84} $\pm$ \num{0.02} (4) \\
         & $\Rey_R \geqslant 4000$ & \num{0.87} $\pm$ \num{0.04} (3) & \num{0.94} $\pm$ \num{0.29} (3) & \num{0.93} $\pm$ \num{0.15} (3) & \num{0.86} $\pm$ \num{0.03} (3) \\
        $f_w = -1$ & all & \num{0.82} $\pm$ \num{0.02} (9) & \num{0.82} $\pm$ \num{0.03} (9) & \num{1.06} $\pm$ \num{0.01} (9) & \num{0.78} $\pm$ \num{0.02} (9) \\
         & $\Rey_R \geqslant 4000$ & \num{0.88} $\pm$ \num{0.03} (5) & \num{0.92} $\pm$ \num{0.09} (5) & \num{1.04} $\pm$ \num{0.03} (5) & \num{0.86} $\pm$ \num{0.04} (5) \\
        $f_w = -2$ & all & \num{0.83} $\pm$ \num{0.07} (4) & \num{0.88} $\pm$ \num{0.08} (4) & \num{1.05} $\pm$ \num{0.05} (4) & \num{0.73} $\pm$ \num{0.09} (4) \\
         & $\Rey_R \geqslant 4000$ & \num{0.86} $\pm$ \num{0.10} (3) & \num{0.91} $\pm$ \num{0.13} (3) & \num{1.06} $\pm$ \num{0.15} (3) & \num{0.77} $\pm$ \num{0.15} (3) \\
        $\hat{x}_\mathrm{ax}/\hat{R} = 2$, $f_w = -1$ & all & \num{0.86} $\pm$ \num{0.05} (4) & \num{0.94} $\pm$ \num{0.08} (4) & \num{1.06} $\pm$ \num{0.06} (4) & \num{0.84} $\pm$ \num{0.06} (4) \\
         & $\Rey_R \geqslant 4000$ & \num{0.89} $\pm$ \num{0.07} (3) & \num{0.98} $\pm$ \num{0.13} (3) & \num{1.06} $\pm$ \num{0.10} (3) & \num{0.88} $\pm$ \num{0.09} (3) \\
        \SI{20}{\percent} \ce{CH4}, $f_w = -1$ & all & \num{0.79} (2) & \num{0.87} (2) & \num{1.04} (2) & \num{0.77} (2) \\
        $\hat{x}_\mathrm{ax}/\hat{R} = 0.5$, $f_w = -0.5$ & all & \num{0.80} $\pm$ \num{0.03} (4) & \num{0.82} $\pm$ \num{0.21} (4) & \num{1.04} $\pm$ \num{0.03} (4) & \num{0.75} $\pm$ \num{0.04} (4) \\
         & $\Rey_R \geqslant 4000$ & \num{0.83} $\pm$ \num{0.06} (3) & \num{0.85} $\pm$ \num{0.43} (3) & \num{1.04} $\pm$ \num{0.04} (3) & \num{0.79} $\pm$ \num{0.06} (3) \\
        $\hat{x}_\mathrm{ax}/\hat{R} = 0.5$, $f_w = -1$ & all & \num{0.81} $\pm$ \num{0.03} (6) & \num{0.76} $\pm$ \num{0.07} (6) & \num{1.04} $\pm$ \num{0.01} (6) & \num{0.73} $\pm$ \num{0.04} (6) \\
         & $\Rey_R \geqslant 4000$ & \num{0.86} $\pm$ \num{0.04} (4) & \num{0.86} $\pm$ \num{0.15} (4) & \num{1.06} $\pm$ \num{0.02} (4) & \num{0.81} $\pm$ \num{0.07} (4) \\
        $\hat{x}_\mathrm{ax}/\hat{R} = 0.5$, $f_w = -2$ & all & \num{0.80} $\pm$ \num{0.07} (4) & \num{0.75} $\pm$ \num{0.11} (4) & \num{1.01} $\pm$ \num{0.03} (4) & \num{0.65} $\pm$ \num{0.12} (4) \\
         & $\Rey_R \geqslant 4000$ & \num{0.83} $\pm$ \num{0.10} (3) & \num{0.79} $\pm$ \num{0.16} (3) & \num{1.02} $\pm$ \num{0.09} (3) & \num{0.70} $\pm$ \num{0.17} (3) \\
        \midrule
        \multicolumn{6}{l}{\textit{Two terms}} \\
        $f_w = -0.5$ & all & \num{1.91} $\pm$ \num{0.24} (4) & -- & \num{1.79} (2) & \num{1.89} $\pm$ \num{0.13} (4) \\
         & $\Rey_R \geqslant 4000$ & \num{1.93} $\pm$ \num{0.54} (3) & -- & -- & \num{1.91} $\pm$ \num{0.26} (3) \\
        $f_w = -1$ & all & \num{1.87} $\pm$ \num{0.22} (5) & \num{1.89} $\pm$ \num{0.08} (7) & \num{2.00} $\pm$ \num{0.04} (8) & \num{1.81} $\pm$ \num{0.07} (9) \\
         & $\Rey_R \geqslant 4000$ & -- & \num{2.01} $\pm$ \num{0.72} (3) & \num{1.82} $\pm$ \num{0.18} (4) & \num{1.90} $\pm$ \num{0.28} (5) \\
        $f_w = -2$ & all & -- & \num{1.93} $\pm$ \num{0.23} (4) & \num{2.00} $\pm$ \num{0.21} (3) & \num{1.79} $\pm$ \num{0.29} (4) \\
         & $\Rey_R \geqslant 4000$ & -- & \num{1.95} $\pm$ \num{0.45} (3) & \num{1.86} (2) & \num{1.82} $\pm$ \num{0.58} (3) \\
        $\hat{x}_\mathrm{ax}/\hat{R} = 2$, $f_w = -1$ & all & -- & \num{2.17} (2) & -- & \num{2.15} (2) \\
        \SI{20}{\percent} \ce{CH4}, $f_w = -1$ & all & \num{1.93} (2) & \num{1.95} (2) & \num{1.84} (2) & \num{1.86} (2) \\
        $\hat{x}_\mathrm{ax}/\hat{R} = 0.5$, $f_w = -0.5$ & all & \num{1.88} $\pm$ \num{0.19} (4) & -- & \num{1.42} $\pm$ \num{0.27} (3) & \num{1.86} $\pm$ \num{0.12} (4) \\
         & $\Rey_R \geqslant 4000$ & \num{1.90} $\pm$ \num{0.42} (3) & -- & \num{1.42} $\pm$ \num{0.27} (3) & \num{1.88} $\pm$ \num{0.26} (3) \\
        $\hat{x}_\mathrm{ax}/\hat{R} = 0.5$, $f_w = -1$ & all & \num{1.85} $\pm$ \num{0.21} (5) & \num{1.85} $\pm$ \num{0.23} (4) & \num{1.70} $\pm$ \num{0.08} (5) & \num{1.82} $\pm$ \num{0.13} (6) \\
         & $\Rey_R \geqslant 4000$ & \num{1.91} $\pm$ \num{0.81} (3) & \num{1.95} (2) & \num{1.42} $\pm$ \num{0.10} (3) & \num{1.89} $\pm$ \num{0.35} (4) \\
        $\hat{x}_\mathrm{ax}/\hat{R} = 0.5$, $f_w = -2$ & all & -- & \num{1.82} $\pm$ \num{0.23} (4) & \num{1.87} $\pm$ \num{0.14} (3) & \num{1.72} $\pm$ \num{0.36} (3) \\
         & $\Rey_R \geqslant 4000$ & -- & \num{1.86} $\pm$ \num{0.39} (3) & \num{1.67} (2) & \num{1.77} (2) \\
        \bottomrule
    \end{tabular}
\end{table}

\textbf{Calibration of the error correlation.} \Cref{tab:calibration} lists the ratios $|e_{2,q}|/(\varepsilon\kappa_q)^2$ of each family that enter the calibration of $C$ and the held-out tests.
\begin{table}[ht!]
    \caption{\label{tab:calibration} Ratio $\bm{|e_2|/(\varepsilon\kappa_q)^2}$ of the two-term error of the numerical solutions to the square of the relative first-order correction, for the values with $\bm{N_u \leqslant |e_2|/3}$. Central values, with $\bm{n}$ the number of computations, and largest upper end $\bm{(|e_2| + N_u)/(\varepsilon\kappa_q)^2}$. Last column, largest ratio of the error of the family to the prediction of the correlation $\bm{E_{2,q} = C (\varepsilon\kappa_q)^2}$ of the paper with $\bm{C}$ refitted without that family, which is the calibrated value for $\bm{\hat{x}_\mathrm{ax}/\hat{R} = 0.5}$.}
    \centering
    \small
    \setlength{\tabcolsep}{4pt}
    \begin{tabular}{llrlrlll}
        \toprule
        Family & Role & $n$ & Ratio, $y_f$ & $n$ & Ratio, $\chi_s$ & Upper end & Left out \\
        \midrule
        $f_w = -0.5$ & calibration & 4 & \num{0.78} to \num{0.83} & 2 & \num{0.98} to \num{1.05} & \num{1.15} & \num{0.84} \\
        $f_w = -1$ & calibration & 5 & \num{0.46} to \num{0.52} & 8 & \num{0.84} to \num{1.12} & \num{1.25} & \num{0.98} \\
        $f_w = -2$ & calibration & 0 & -- & 3 & \num{0.89} to \num{0.92} & \num{1.04} & \num{0.73} \\
        $\hat{x}_\mathrm{ax}/\hat{R} = 2$, $f_w = -1$ & held out & 0 & -- & 1 & \num{0.78} & \num{0.99} & \num{0.63} \\
        \SI{20}{\percent} \ce{CH4}, $f_w = -1$ & held out & 2 & \num{0.67} to \num{0.69} & 2 & \num{0.97} to \num{1.02} & \num{1.08} & \num{0.82} \\
        $\hat{x}_\mathrm{ax}/\hat{R} = 0.5$, $f_w = -0.5$ & held out & 4 & \num{2.18} to \num{2.40} & 3 & \num{1.38} to \num{1.88} & \num{2.80} & \num{1.92} \\
        $\hat{x}_\mathrm{ax}/\hat{R} = 0.5$, $f_w = -1$ & held out & 5 & \num{1.07} to \num{1.25} & 5 & \num{1.00} to \num{1.74} & \num{1.84} & \num{1.39} \\
        $\hat{x}_\mathrm{ax}/\hat{R} = 0.5$, $f_w = -2$ & held out & 1 & \num{0.95} & 3 & \num{0.99} to \num{1.16} & \num{1.29} & \num{0.92} \\
        \bottomrule
    \end{tabular}
\end{table}

\section{Accuracy of the Two-Term Expansion}
\begin{figure}[ht!]
    \centering
    \includegraphics{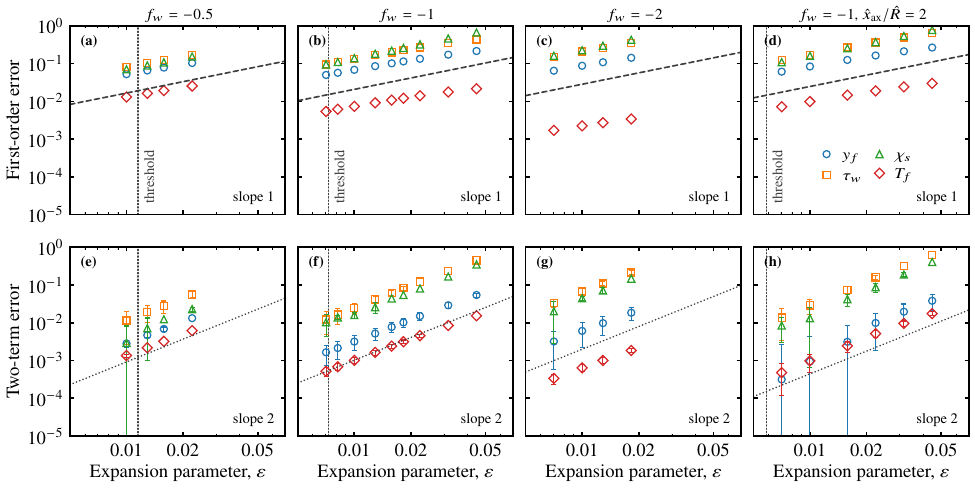}
    \caption{Relative error of the model against the numerical solutions as a function of $\bm{\varepsilon = \Rey_R^{-1/2}}$. (a)--(d)~First-order model; (e)--(h)~two-term expansion. Columns: $\bm{f_w = -0.5}$, $\bm{-1}$ and $\bm{-2}$ for $\bm{\hat{x}_\mathrm{ax}/\hat{R} = 1}$, and $\bm{f_w = -1}$ for $\bm{\hat{x}_\mathrm{ax}/\hat{R} = 2}$; symbols: errors of the flame stand-off distance, $\bm{y_f}$ (circles), wall shear stress, $\bm{\tau_w}$ (squares), stoichiometric scalar dissipation rate, $\bm{\chi_s}$ (triangles), and flame temperature, $\bm{T_f}$ (diamonds), relative to the computed values; error bars: numerical uncertainty $\bm{N_u}$ where it exceeds a tenth of the two-term error; dashed and dotted lines: slopes 1 and 2; vertical dotted lines: first-order adequacy threshold of the manuscript, with the computations above it to its left. For $\bm{f_w = -2}$, the threshold, $\bm{\Rey_R = 5.2 \times 10^4}$, lies outside the plotted range.}
    \label{fig:s_cfd_error}
\end{figure}

\label[section]{sec:s_accuracy}
\textbf{Errors against the Navier-Stokes solutions.} The errors relative to the computed values are shown in \cref{fig:s_cfd_error}. For the families of \cref{fig:s_cfd_error}, the first-order errors decrease approximately as $\varepsilon$, whereas the two-term errors decrease approximately as $\varepsilon^2$, as expected from the expansion. The scalar dissipation rate of the shorter porous section, which is not shown, decreases more slowly (\cref{tab:slopes}). The exponents of the weighted fits are given in \cref{tab:slopes}. However, these exponents are only suggestive, since few computations lie above the first-order threshold. From $\Rey_R = 1000$ upward, the two-term errors of all four quantities are smaller than the first-order ones. At $\Rey_R = 500$, the two-term error in wall shear is comparable to the first-order error, which marks the lower end of the expansion for $f_w = -1$. This is consistent with the overprediction of the flame shift and the reduced velocity gradient above the layer in the stagnation-line profiles of the manuscript at this Reynolds number. For $\hat{x}_\mathrm{ax}/\hat{R} = 1$ and $f_w = -1$, the second-order terms reduce the first-order errors severalfold in the range of laboratory burners, and the reduction is larger for the stand-off distance than for the scalar dissipation rate. The errors are smaller for weaker injection. For the longer porous section at low Reynolds numbers, the two-term errors in the scalar dissipation rate remain large, which is consistent with the error correlation.

\textbf{Displacement coefficient from the numerical solutions.} The slip-velocity gradient extracted above the layer contains outer terms of higher order, which are large for variable density. For $\hat{x}_\mathrm{ax}/\hat{R} = 1$ and $f_w = -1$, its extrapolations bracket the march value, but they cannot discriminate it. For $\hat{x}_\mathrm{ax}/\hat{R} = 2$, the estimates from the slip velocity and from the wall pressure approach the march value as $O(\varepsilon)$, and their extrapolations through the highest Reynolds numbers bracket it, with a spread that depends on the fit range. Hence, the solutions are consistent with the march value within about \SI{10}{\percent}.

\textbf{Laminar flow.} The solutions verify the laminar model, and they do not show that the physical flow is laminar. For $\Rey_R \geqslant 6000$, the displacement-thickness Reynolds number exceeds 520, the critical value of the Blasius layer \cite{schlichting_boundarylayer_2017}, between about 7 and 19 nose radii downstream. This value is only a reference for a heated, injected, and inflectional layer. Since $U_{2,1}$ depends on the whole layer, downstream changes in the displaced flux affect it through the weighted integral in the manuscript (\cref{sec:s_march}). For $f_w = -1$ and $\hat{x}_\mathrm{ax}/\hat{R} = 1$, a displacement thickness that grows as in a turbulent layer beyond 7 nose radii increases the magnitude of $U_{2,1}$ by about \SI{40}{\percent}, and the change is smaller when this growth starts further downstream. At $\Rey_R = 6000$, the resulting change in wall shear is comparable to its two-term error. This is a prescribed perturbation, which does not predict transition, and it also differs from the numerical convergence of the march.

\begin{figure}[ht!]
    \centering
    \includegraphics{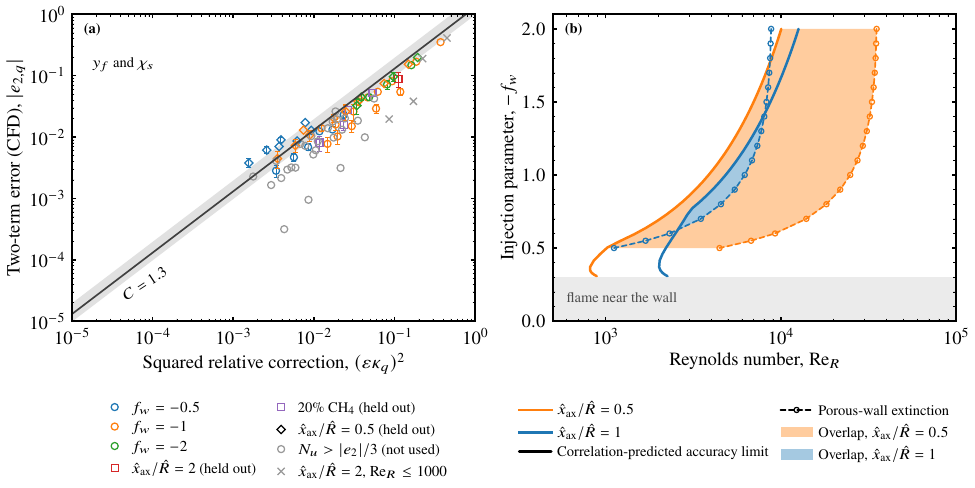}
    \caption{Accuracy of the two-term expansion. (a)~Relative two-term error of the numerical solutions, $\bm{|e_{2,q}|}$, versus the square of the relative first-order correction, $\bm{(\varepsilon\kappa_q)^2}$, for the flame stand-off distance and the scalar dissipation rate; circles: calibration families; squares: held-out families; diamonds: held-out families with $\bm{\hat{x}_\mathrm{ax}/\hat{R} = 0.5}$ in the color of each $\bm{f_w}$; grey circles: values with $\bm{N_u > |e_{2,q}|/3}$, not used; grey crosses: $\bm{\hat{x}_\mathrm{ax}/\hat{R} = 2}$ at $\bm{\Rey_R \leqslant 1000}$, outside the calibration domain; error bars: numerical uncertainty $\bm{N_u}$ where it exceeds a tenth of the error; line: $\bm{C = 1.3}$; grey band: $\bm{C \times [0.75, 1.5]}$. (b)~Smallest Reynolds number for which the error correlation predicts errors within the tolerances, which is optimistic for $\bm{\hat{x}_\mathrm{ax}/\hat{R} = 0.5}$ at weak injection (see (a)), for $\bm{\hat{x}_\mathrm{ax}/\hat{R} = 0.5}$ and 1; grey region: flame within about one viscous length of the wall; lines with symbols: extinction of the finite-rate porous-wall flame for $\bm{\hat{R}} = \SI{50.8}{\milli\meter}$ and \SI{25.4}{\milli\meter}, i.e., a porous section with an axial length of \SI{25.4}{\milli\meter}; light orange and light blue regions: nominal candidate overlaps for $\bm{\hat{x}_\mathrm{ax}/\hat{R} = 0.5}$ and 1, respectively, and $\bm{C = 1.3}$, between the two-term accuracy limit and this planar extinction.}
    \label{fig:s_applicability}
\end{figure}

\textbf{Calibration and held-out tests.} The constant of the error correlation of the manuscript, $E_{2,q} = C (\varepsilon \kappa_q)^2$, is calibrated on the families with $\hat{x}_\mathrm{ax}/\hat{R} = 1$ and the reference mixture (\cref{fig:s_applicability}a and \cref{tab:calibration}). The ratio $|e_{2,q}|/(\varepsilon \kappa_q)^2$, with $e_{2,q}$ relative to the computed value and $\kappa_q$ to $q_1$, is of order one for all calibration values, and it stays below $C = \num{1.3}$ when the numerical uncertainty is included. Therefore, this value is adopted. The ratios of the held-out families, i.e., $\hat{x}_\mathrm{ax}/\hat{R} = 2$ and \SI{20}{\percent} \ce{CH4}, remain below $C$. Refitting $C$ with each family left out changes it only slightly, and the left-out errors stay below the prediction. However, the held-out set is small, with few resolved values for the longer injection length and the diluted fuel.

The shorter porous section, $\hat{x}_\mathrm{ax}/\hat{R} = \num{0.5}$, i.e., the same porous section on a nose radius twice as large, is a further held-out test with three injection rates (\cref{fig:s_applicability}a, diamonds). Its ratios remain below $C$ for $f_w = -2$. For $f_w = -1$, only the scalar dissipation rate at the highest Reynolds number exceeds $C$, and its error does not decrease further. For $f_w = -0.5$, all included values exceed $C$, by up to a factor of about two. For this weak injection, however, the first-order corrections are small, and the two-term errors remain within a few percent above $\Rey_R = 2000$. Therefore, the correlation underestimates the already small two-term errors of this family. These tests support the correlation with $C = \num{1.3}$ for the calibration families, longer injection length, diluted fuel, and shorter injection length at $f_w = -2$ only. Outside these cases, the correlation is not a bound.

\textbf{Accuracy limits and extinction.} For $\hat{x}_\mathrm{ax}/\hat{R} = 1$, the correlation places the two-term accuracy limit at Reynolds numbers of a few thousand, and the limit rises with injection (\cref{fig:s_applicability}b). These limits are proportional to $C$. The longer porous section raises the limits, and the first-order threshold of the manuscript lies well above them for all extents. The upper bound of a burning range is estimated with the planar finite-rate porous-wall flame (\cref{sec:s_porous}), whose extinction strain increases with injection. For weak injection, the flame does not burn at the lowest strain rates considered. This extinction depends on the chemical time, and hence its Reynolds number, $\hat{a}_\mathrm{ext} \hat{R}^2/\hat{\nu}_\infty$, depends on body size. In \cref{fig:s_applicability}b, it is shown for a porous section with an axial length of \SI{25.4}{\milli\meter}, i.e., for $\hat{R} = \SI{25.4}{\milli\meter}$ and \SI{50.8}{\milli\meter} at $\hat{x}_\mathrm{ax}/\hat{R} = 1$ and \num{0.5}, respectively. For $\hat{x}_\mathrm{ax}/\hat{R} = 1$, the accuracy limit and the planar extinction give a nominal candidate overlap only at intermediate injection rates. Since the planar extinction includes neither curvature nor displacement, and the overlap ends depend on interpolation between the sampled injection rates, these ends are approximate. The extinction Reynolds number exceeds the accuracy limit by less than half. Therefore, the overlap vanishes within the band $C \times [0.75, 1.5]$, and it is not an operating envelope of the cylinder. The same screening yields no overlap for the weakest injection, the longer porous section, or diluted fuels, whose porous-wall flames extinguish at low strain rates (\cref{tab:dilution}). This result does not show that a flame cannot burn there.

The same porous section on a nose radius twice as large, i.e., $\hat{x}_\mathrm{ax}/\hat{R} = \num{0.5}$, gives a different picture. Its displacement coefficient is smaller, which lowers the accuracy limits, and the larger radius raises the extinction Reynolds number fourfold at the same strain rate. As a result, the extinction Reynolds number is several times the accuracy limit, and the nominal overlap covers all sampled injection rates. This overlap vanishes only for values of $C$ several times larger than the calibrated one. The first-order threshold also lies below the extinction except for the strongest injection. Without the correlation, the numerical solutions are consistent with the accuracy side of this overlap at the sampled points. For every computed $f_w$, the two-term errors fall within tolerance at Reynolds numbers below the extinction value, including weak injection, where the correlation is optimistic. However, these errors rely on the transferred boundary allowance and are provisional.
\begin{table}[t!]
    \caption{\label{tab:dilution} Effect of fuel dilution on the two-term accuracy limit (variable density). $\bm{\Rey_\mathrm{min}}$, smallest Reynolds number with an error estimate from the correlation of the paper, with $\bm{C = 1.3}$, below \SI{2}{\percent} for the stand-off and \SI{5}{\percent} for the scalar dissipation, rounded to two digits ($\bm{\Rey_\mathrm{min} \propto C}$); $\bm{\Rey_\mathrm{ext}}$, extinction of the finite-rate porous-wall flame for $\bm{\hat{R}} = \SI{25.4}{\milli\meter}$ ($\bm{\Rey_\mathrm{ext}}$ scales with $\bm{\hat{R}^2}$); first-order stand-off distance in viscous lengths, $\bm{y_{f1}/\varepsilon}$, and relative first-order correction of the stand-off distance, $\bm{|\varepsilon\kappa_{y_f}|}$, at $\bm{\Rey_R = 4000}$ and $\bm{\hat{x}_\mathrm{ax}/\hat{R} = 1}$; --, no detached flame or no burning solution.}
    \centering
    \small
    \setlength{\tabcolsep}{4pt}
    \begin{tabular}{lrrrrrrrr}
        \toprule
        Fuel & $Z_s$ & $\gamma$ & $f_w$ & $\Rey_\mathrm{min}$, $\hat{x}_\mathrm{ax}/\hat{R} = 1$ & $\Rey_\mathrm{min}$, $\hat{x}_\mathrm{ax}/\hat{R} = 2$ & $\Rey_\mathrm{ext}$ & $y_{f1}/\varepsilon$ & $|\varepsilon\kappa_{y_f}|$ \\
        \midrule
        \SI{30}{\percent} \ce{CH4} & \num{0.228} & \num{5.76} & \num{-0.5} & \num{2200} & \num{3700} & \num{1100} & \num{2.85} & \num{0.092} \\
         &  &  & \num{-1.0} & \num{4800} & \num{5800} & \num{6200} & \num{5.56} & \num{0.122} \\
        \SI{20}{\percent} \ce{CH4} & \num{0.317} & \num{5.21} & \num{-0.5} & \num{2200} & \num{4100} & -- & \num{1.61} & \num{0.092} \\
         &  &  & \num{-1.0} & \num{3100} & \num{4900} & \num{1900} & \num{4.20} & \num{0.108} \\
        \SI{15}{\percent} \ce{CH4} & \num{0.388} & \num{4.76} & \num{-0.5} & \num{3000} & \num{5800} & -- & \num{0.91} & \num{0.108} \\
         &  &  & \num{-1.0} & \num{2700} & \num{4600} & \num{440} & \num{3.32} & \num{0.102} \\
        \bottomrule
    \end{tabular}
\end{table}

\textbf{Slot comparison.} The slot flame of the manuscript has a gap of \SI{11.8}{\milli\meter}, both streams at \SI{300}{\kelvin} and atmospheric pressure, and inlet velocities of \SI{29.9}{\centi\meter\per\second} (fuel) and \SI{33.4}{\centi\meter\per\second} (oxidizer). Their ratio is held fixed while both are scaled to the nominal strain rate. Each manuscript profile is plotted against the distance from its stagnation plane, where the normal velocity vanishes, and the hatched region marks the wall of the first-order parabola. This comparison is contextual, since the nominal strain rate does not match the fuel flux or the local strain at the flame. For the parabola of the slot comparison in the manuscript, the first-order corrections are about a tenth of the stand-off distance and about a fifth of the wall shear and the scalar dissipation rate. The correlation estimates two-term errors of a few percent for these quantities. The finite-rate porous-wall flame has a lower temperature rise than the flame sheet, and its differences from the flame sheet grow with injection at the same strain rate (\cref{tab:porous}). These differences are larger than the two-term truncation error of the stand-off distance. Therefore, the two-term expansion is accurate for the flame-sheet problem, whereas the flame-sheet model itself differs more from the detailed formulation.

\section{Convergence of the Reacting March}
\begin{figure}[ht!]
    \centering
    \includegraphics{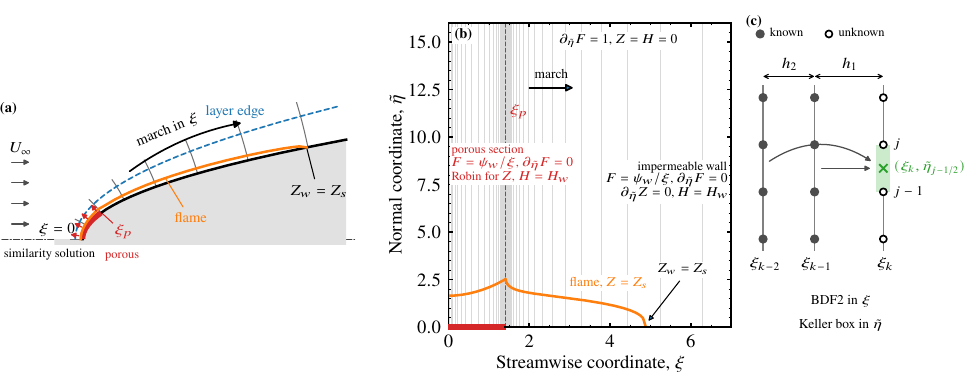}
    \caption{Schematic of the boundary-layer march. (a)~Physical plane, with the thickness of the layer exaggerated and the lines of constant $\bm{\xi}$ through the layer; (b)~computational plane, with the flame sheet $\bm{Z = Z_s}$ of the reacting march for $\bm{f_w = -1}$ and $\bm{\hat{x}_\mathrm{ax}/\hat{R} = 1}$ and the wall and edge conditions, every 12th station shown; (c)~stencil of the second-order backward difference in $\bm{\xi}$ and the Keller box in $\bm{\tilde{\eta}}$, centered at $\bm{(\xi_k, \tilde{\eta}_{j-1/2})}$. Porous section in brown.}
    \label{fig:s_march}
\end{figure}

\label[section]{sec:s_march}
The reacting march uses the Keller box scheme in $\tilde{\eta}$ and a second-order backward difference in $\xi$, with steps that grow in proportion to $\xi$ away from the nose and are refined near $\xi_p$. The march is sketched in \cref{fig:s_march}. It starts from the similarity solution at $\xi = 0$, and Newton iterations solve each station given the two previous stations. The flame is not tracked, and it reaches the wall where $Z_w = Z_s$ downstream of the porous section.
\Cref{tab:march} lists the changes in $U_{2,1}$ for the reference mixture with the march step, the normal resolution, the domain height, the downstream cut-off, and the contributions of the four body parts. The nose region gives most of $U_{2,1}$. The thinning of the hot layer downstream of the porous section contributes positively, whereas the layer beyond $\xi = 12$ contributes a small part.
\begin{table}[ht!]
    \caption{\label{tab:march} Reacting march for the reference mixture. Baseline $\bm{U_{2,1}}$ (step ratio 0.02, normal grid of 401 nodes on $\bm{\tilde{\eta} \leqslant 16}$, or 601 on $\bm{\tilde{\eta} \leqslant 24}$ for $\bm{f_w = -2}$ and $\bm{\hat{x}_\mathrm{ax}/\hat{R} = 2}$, $\bm{\xi_{\max} = 1000}$); largest change for step ratios 0.04 and 0.01, for half and twice the normal nodes, for a normal domain height larger by 8, and for the cut-off $\bm{\xi_{\max} = 100}$; contributions to $\bm{U_{2,1}}$ of the nose region, $\bm{\xi < 0.8\,\xi_p}$, the porous-edge neighborhood, $\bm{0.8\,\xi_p < \xi < 1.25\,\xi_p}$, the layer downstream up to the outlet of the numerical solutions, $\bm{\xi < 12}$, and the layer beyond it.}
    \centering
    \small
    \setlength{\tabcolsep}{4pt}
    \begin{tabular}{rr r rrrr rrrr}
        \toprule
         & & & \multicolumn{4}{c}{Change} & \multicolumn{4}{c}{Contribution} \\
        \cmidrule(lr){4-7} \cmidrule(lr){8-11}
        $\hat{x}_\mathrm{ax}/\hat{R}$ & $f_w$ & $U_{2,1}$ & Step & Nodes & Height & Cut-off & Nose & Edge & $\xi < 12$ & $\xi > 12$ \\
        \midrule
        1 & \num{-0.5} & \num{-1.909} & \num{4.0e-04} & \num{2.7e-03} & $<$\num{1e-6} & \num{1.9e-06} & \num{-3.39} & \num{-0.43} & \num{1.89} & \num{0.02} \\
        1 & \num{-1} & \num{-2.611} & \num{4.7e-04} & \num{3.3e-03} & $<$\num{1e-6} & \num{2.9e-06} & \num{-4.14} & \num{-0.21} & \num{1.65} & \num{0.09} \\
        1 & \num{-2} & \num{-3.988} & \num{3.7e-03} & \num{3.2e-03} & $<$\num{1e-6} & \num{6.7e-07} & \num{-6.24} & \num{0.53} & \num{1.42} & \num{0.30} \\
        2 & \num{-0.5} & \num{-3.398} & \num{3.1e-04} & \num{2.0e-03} & $<$\num{1e-6} & \num{3.7e-06} & \num{-4.33} & \num{-0.51} & \num{1.38} & \num{0.07} \\
        2 & \num{-1} & \num{-4.416} & \num{1.6e-03} & \num{3.3e-03} & $<$\num{1e-6} & \num{3.2e-06} & \num{-5.45} & \num{-0.26} & \num{1.07} & \num{0.22} \\
        2 & \num{-2} & \num{-6.560} & \num{1.5e-02} & \num{4.2e-03} & $<$\num{1e-6} & \num{5.3e-06} & \num{-8.51} & \num{0.44} & \num{1.02} & \num{0.50} \\
        0.5 & \num{-0.5} & \num{-0.264} & \num{7.5e-04} & \num{3.8e-03} & $<$\num{1e-6} & \num{7.2e-07} & \num{-2.56} & \num{-0.20} & \num{2.49} & \num{0.01} \\
        0.5 & \num{-1} & \num{-0.624} & \num{1.2e-03} & \num{3.8e-03} & $<$\num{1e-6} & \num{1.9e-06} & \num{-3.09} & \num{0.00} & \num{2.42} & \num{0.04} \\
        0.5 & \num{-2} & \num{-1.398} & \num{3.6e-03} & \num{2.6e-03} & $<$\num{1e-6} & \num{1.9e-06} & \num{-4.52} & \num{0.82} & \num{2.16} & \num{0.13} \\
        \bottomrule
    \end{tabular}
\end{table}

A change in the displaced flux beyond a station $\xi_t$ changes $U_{2,1}$ according to the weighted integral of the manuscript, with weight $4/(\pi \xi^2)$. The stations $s_t = 7$ and 19 nose radii correspond to $\xi_t = 3.4$ and 5.9. Two perturbations of the displacement thickness beyond $s_t$ are considered, a doubling and a growth proportional to $(s/s_t)^{0.3}$, which turns the laminar growth, $s^{1/2}$, into that of a turbulent layer, $s^{0.8}$. For $\hat{x}_\mathrm{ax}/\hat{R} = 1$ and $s_t = 7$, both perturbations increase the magnitude of $U_{2,1}$ substantially. The doubling has the larger effect, and its effect grows with injection. Turbulent growth increases the magnitude by about \SI{40}{\percent} at all injection rates. For $s_t = 19$, both changes are roughly halved. These values are perturbation estimates, and they do not result from transition calculations.

\section{Finite-Rate Porous-Wall Flame}\label[section]{sec:s_porous}
The finite-rate porous-wall flame is the planar stagnation flow in the manuscript, with the porous wall as a no-slip fuel inlet at \SI{300}{\kelvin}, species-flux conditions of the porous wall, and the oxidizer entering as the Hiemenz flow at the prescribed strain rate. The extinction strain is a refined numerical estimate of the upper burning-branch extinction strain. The prescribed control strain is raised from a burning state until the burning solution is lost. Each cold state is rematched to the prescribed strain, since its eigenvalue differs noticeably from it. The bracket between the last burning and the first cold control strain is the search resolution, and it is not the total uncertainty of the estimate. The search does not compute the unstable branch. A finer grid and a wider domain change the estimate by less than \SI{1}{\percent}. \Cref{tab:porous} lists the estimates and selected comparisons with the first-order flame sheet. The detailed flame is located where the Bilger mixture fraction, normalized with the supplied fuel and oxidizer streams, equals $Z_s$.
\begin{table}[ht!]
\caption{\label{tab:porous} Finite-rate porous-wall flame. (a)~Refined numerical estimate of the upper burning-branch extinction strain, i.e., last burning control strain, first cold control strain after rematching, change on a finer grid, and peak temperature rise, $\bm{T_\mathrm{max}}$, at the last burning state. (b)~Flame position from the wall in viscous lengths, $\bm{y_f/\varepsilon = \hat{y}_f (\hat{a}/\hat{\nu}_\infty)^{1/2}}$, peak temperature rise and wall mixture fraction of the first-order flame sheet and of the detailed solution, located where the Bilger mixture fraction normalized with the supplied streams equals $\bm{Z_s}$.}
\centering
\small
\begin{tabular}{llrrrr}
    \toprule
    \multicolumn{6}{l}{(a)~Extinction} \\
    \midrule
    Fuel & $f_w$ & Burning (\si{\per\second}) & Cold (\si{\per\second}) & Finer grid (\si{\percent}) & $T_\mathrm{max}$ \\
    \midrule
    \SI{30}{\percent} & \num{-0.3} & \multicolumn{4}{l}{no burning solution found for $\hat{a} \geqslant \SI{10}{\per\second}$} \\
    \SI{30}{\percent} & \num{-0.4} & \multicolumn{4}{l}{no burning solution found for $\hat{a} \geqslant \SI{10}{\per\second}$} \\
    \SI{30}{\percent} & \num{-0.5} & \num{27.5} & \num{27.6} & \num{+0.17} & \num{3.75} \\
    \SI{30}{\percent} & \num{-0.55} & \num{41.7} & \num{41.8} & \num{-0.08} & \num{3.84} \\
    \SI{30}{\percent} & \num{-0.6} & \num{57.1} & \num{57.2} & \num{-0.32} & \num{3.88} \\
    \SI{30}{\percent} & \num{-0.7} & \num{86.1} & \num{86.3} & \num{-0.08} & \num{4.00} \\
    \SI{30}{\percent} & \num{-0.8} & \num{112.2} & \num{112.4} & \num{-0.08} & \num{4.09} \\
    \SI{30}{\percent} & \num{-0.9} & \num{134.5} & \num{134.8} & \num{+0.17} & \num{4.16} \\
    \SI{30}{\percent} & \num{-1} & \num{153.3} & \num{153.6} & \num{-0.08} & \num{4.21} \\
    \SI{30}{\percent} & \num{-1.1} & \num{168.7} & \num{169.1} & \num{-0.08} & \num{4.24} \\
    \SI{30}{\percent} & \num{-1.2} & \num{181.0} & \num{181.4} & \num{-0.08} & \num{4.29} \\
    \SI{30}{\percent} & \num{-1.3} & \num{190.7} & \num{191.1} & \num{-0.08} & \num{4.29} \\
    \SI{30}{\percent} & \num{-1.4} & \num{198.3} & \num{198.8} & \num{+0.17} & \num{4.32} \\
    \SI{30}{\percent} & \num{-1.5} & \num{204.1} & \num{204.5} & \num{+0.17} & \num{4.35} \\
    \SI{30}{\percent} & \num{-1.6} & \num{208.6} & \num{209.0} & \num{+0.17} & \num{4.36} \\
    \SI{30}{\percent} & \num{-1.7} & \num{211.8} & \num{212.2} & \num{+0.17} & \num{4.37} \\
    \SI{30}{\percent} & \num{-1.8} & \num{214.1} & \num{214.6} & \num{+0.17} & \num{4.38} \\
    \SI{30}{\percent} & \num{-1.9} & \num{215.5} & \num{216.0} & \num{+0.17} & \num{4.39} \\
    \SI{30}{\percent} & \num{-2} & \num{216.9} & \num{217.4} & \num{-0.08} & \num{4.36} \\
    \SI{20}{\percent} & \num{-0.5} & \multicolumn{4}{l}{no burning solution found for $\hat{a} \geqslant \SI{10}{\per\second}$} \\
    \SI{20}{\percent} & \num{-1} & \num{47.3} & \num{47.4} & \num{+0.42} & \num{3.85} \\
    \SI{20}{\percent} & \num{-1.5} & \num{87.3} & \num{87.5} & \num{+0.17} & \num{4.04} \\
    \SI{15}{\percent} & \num{-0.5} & \multicolumn{4}{l}{no burning solution found for $\hat{a} \geqslant \SI{10}{\per\second}$} \\
    \SI{15}{\percent} & \num{-1} & \num{10.9} & \num{11.0} & \num{+0.67} & \num{3.57} \\
    \bottomrule
\end{tabular}
\par\medskip
\begin{tabular}{rrrrrrrrr}
    \toprule
    \multicolumn{9}{l}{(b)~Detailed solution and flame sheet, \SI{30}{\percent} \ce{CH4}} \\
    \midrule
     & & \multicolumn{3}{c}{$y_f/\varepsilon$} & \multicolumn{2}{c}{$T_\mathrm{max}$} & \multicolumn{2}{c}{$Z_w$} \\
    \cmidrule(lr){3-5} \cmidrule(lr){6-7} \cmidrule(lr){8-9}
    $f_w$ & $\hat{a}$ (\si{\per\second}) & Sheet & Detailed & Diff. (\si{\percent}) & Sheet & Detailed & Sheet & Detailed \\
    \midrule
    \num{-0.5} & \num{25} & \num{2.85} & \num{2.89} & \num{+1.4} & \num{3.68} & \num{3.81} & \num{0.450} & \num{0.519} \\
    \num{-0.8} & \num{100} & \num{4.65} & \num{4.24} & \num{-8.9} & \num{4.73} & \num{4.23} & \num{0.622} & \num{0.714} \\
    \num{-1} & \num{50} & \num{5.56} & \num{5.09} & \num{-8.5} & \num{5.08} & \num{4.72} & \num{0.715} & \num{0.808} \\
    \num{-1} & \num{100} & \num{5.56} & \num{5.00} & \num{-10.1} & \num{5.08} & \num{4.53} & \num{0.715} & \num{0.809} \\
    \num{-2} & \num{100} & \num{8.64} & \num{7.47} & \num{-13.5} & \num{5.69} & \num{4.86} & \num{0.963} & \num{0.990} \\
    \bottomrule
\end{tabular}
\end{table}

\section{Codes and Data}\label[section]{sec:s_repro}
The slot and porous-wall flames use the modified Cantera code of \cref{sec:s_slot}. The detailed solutions use the San Diego mechanism \cite{ucsd_san_2016} with its thermodynamic and transport data. The Navier-Stokes solutions use \texttt{bsLowMachFoam}, a steady low-Mach solver built on OpenFOAM 14 \cite{weller_tensorial_1998}, which solves the equations of the manuscript with $\rho (1 + T) = 1$, $\rho \mu = 1$, and a Burke-Schumann flame. The grids follow parabolic coordinates, with a refinement ratio of 2, three levels of \num{15360}, \num{61440}, and \num{245760} cells, and half of the normal cells within $30 \varepsilon$ of the wall. Wall gradients are obtained by polynomial fits in $\eta$ over $0 < \eta < 1$. The finite-rate porous-wall flames and the slot flames are computed with the scripts \texttt{cantera\_porous.py} and \texttt{cantera\_slot.py} with this Cantera version. Numerical uncertainty, error exponents, and calibration are obtained with \texttt{verification.py}, and march convergence with \texttt{march\_convergence.py}. The remaining codes and data, i.e., the asymptotic solver, the solver \texttt{bsLowMachFoam} and its cases, the post-processing scripts and the data, are available from the corresponding author upon request.

\bibliographystyle{new-aiaa}
\bibliography{references}

\end{document}